\documentclass[aps,prb,twocolumn,amsmath,amssymb]{revtex4-1}
\usepackage{graphicx}
\usepackage{enumitem}
\usepackage[colorlinks=true,citecolor=blue]{hyperref}

\usepackage[normalem]{ulem}

\newcommand{\circled}[1]{\raisebox{.5pt}{\textcircled{\raisebox{-.9pt}{#1}}}}

\begin{document}

\title{Relativistic Gurzhi effect in channels of Dirac materials}
\author{Oleksiy Kashuba}
\email[Email: ]{okashuba@physik.uni-wuerzburg.de}
%
\author{Bj\"orn Trauzettel}
\affiliation{Theoretische Physik IV, Institut f\"ur Theoretische Physik und Astrophysik, Universit\"at W\"urzburg, 97074 W\"urzburg, Germany}
\author{Laurens W.\ Molenkamp}
\affiliation{Experimentelle Physik III, Physikalisches Institut, Universit\"at W\"urzburg, 97074 W\"urzburg, Germany}
\begin{abstract}
Charge transport in channel-shaped 2D Dirac systems is studied employing the Boltzmann equation.
The dependence of the resistivity on temperature and chemical potential is investigated.
An accurate understanding of the influence of electron-electron interaction and material disorder allows us to identify a parameter regime, where the system reveals hydrodynamic transport behavior.
We point out the conditions for three Dirac fermion specific features: heat flow hydrodynamics, pseudo\-diffusive transport, and the electron-hole scattering dominated regime.
It is demonstrated that for clean samples the relativistic Gurzhi effect, a definite indicator of hydrodynamic transport, can be observed.
\end{abstract}

\maketitle

\section{Introduction}

The possibility of hydrodynamical transport of charge carriers in condensed matter systems has fascinated the scientific community for a long time.
A system without an underlying lattice, despite particle interaction, experiences no net momentum relaxation, except at the boundaries.
These conditions are auspicious for hydrodynamic transport. 
Solid state systems, on the contrary, tend to dissipate the net momentum of excitations to a phonon bath.
The first experiment with the unambiguous indications of hydrodynamical transport behavior of electrons was done 20 years ago in (Al,Ga)As heterostructures~\cite{Molenkamp1994a,Jong1995}, on the basis of the non\-relativistic Gurzhi effect.
This effect is characterized by a non\-monotonic dependence of the resistivity on the electron gas temperature reaching a maximum at the crossover from the quasi\-ballistic (Knudsen) to the hydrodynamic (Poiseuille) regime~\cite{Gurzhi1963,*Gurzhi1964,*Gurzhi1968}.
The main experimental obstacle to observe the Gurzhi effect is the requirement of high mobility of charge carriers together with strong inter\-carrier scattering, which compels the carriers to propagate as a whole.
Similar conditions, but based on a different mechanism, can be realized in ultra pure metals~\cite{Moll2016}.

Despite of experimental difficulties, solid state systems are highly interesting objects to study, since they offer an opportunity for the investigation of rich physics based on different dimensionalities, complex dispersion relations, topologically non\-trivial band structures, Dirac vs.\ Sch{\"o}dinger fermions, etc.
In this paper, we analyze electron transport in a system composed of Dirac fermions.
The best known example of such a material is graphene, for which a number of experiments reported the observation of fluid\-like behavior~\cite{Bandurin2016,Crossno2016,Morpurgo2017}.
Recent advances in manufacturing topological materials expand the variety of realizations of 2D Dirac Hamiltonians, for instance, based on the surface states of a 3D topological insulator (TI)~\cite{Hancock2011,Buttner2011}.
However, the Gurzhi effect in Dirac materials has not been discussed in the literature so far.
We close this gap and point out the rich transport physics of Dirac fermions in channel geometries, see Fig.~\ref{fig:roadmap} below. 

Dirac materials are remarkable for their two related characteristics: strong spin-momentum locking and a Dirac cone spectrum with both valence and conduction band touching each other.
The strong spin-momentum locking, which plays a crucial role in the quantum spin Hall effect~\cite{Konig2007}, affects the scattering of the 2D surface states only quantitatively~\cite{Kechedzhi2008}.
The massless relativistic spectrum, in contrast, strongly influences thermalization processes, making them similar to those occurring in bad metals~\cite{Fritz2008}.
This spectrum results in a finite resistivity even for absolutely clean systems due to interactions~\cite{Ziegler2007,Kashuba2008}.
Furthermore, the system can experience so-called collision-dominated nonlinear hydrodynamics~\cite{Muller2009a,Briskot2015}.

Here, we study charge transport in a channel of a Dirac material.
The goal is to understand the crossover regimes where hydrodynamic equations for Dirac systems~\cite{Narozhny2015,Lucas2016a,Narozhny2017,Lucas2018} are not yet fully valid, but the system exhibits a tendency to hydrodynamic behavior, the relativistic analog of the Gurzhi effect~\footnote{%
The full Hamiltonian including Coulomb interaction is not relativistic i.e.\ not invariant with respect to Lorentz transformations.
However, the word ``relativistic'' is widely used in the modern literature of condensed matter physics (e.g. on graphene or surface states of 3D topological insulators) referring to the Dirac kinetic part of the Hamiltonian.}.

We exploit the kinetic equation approach, which allows us to take into account different scattering channels and study their interplay.
Our analysis implies interesting future direction of research: 
(i) It allows for a concrete comparison with experiments done on surfaces of 3D TIs.
(ii) Eventually, we aim to connect our predictions with complementary approaches based on the gauge-gravity duality~\cite{Sachdev2017}.

The paper is organized as follows: Section~\ref{sec:roadmap} is devoted to the qualitative description of various transport regimes and contains their illustrative classification according to the system parameters.
Section~\ref{sec:kinetic} depicts a rigorous description of the kinetic equation and collision integrals, and demonstrates how they can be solved.
The solutions and their interpretation indicating the presence of hydrodynamical behavior are presented in Section~\ref{sec:results}.
We summarize in Section~\ref{sec:summary} and move technical details into the Appendices.

\section{Qualitative analysis}
\label{sec:roadmap}

The stationary Boltzmann equation for the quasiparticles of our system under consideration can be written as~\cite{ll10}
\begin{equation}
\mathbf{v} \cdot \partial_\mathbf{r} f + e\mathbf{E}\cdot\partial_\mathbf{p} f = \mathcal{I}[f],
\label{eq:boltzmann}
\end{equation}
where $f\equiv f(\mathbf{r},\breve{\mathbf{p}})$ is a distribution function, and $\breve{\mathbf{p}}$ denotes the complete set of quantum numbers, that unambiguously define the quantum state (for Dirac spectra, for example, it is a band index and the momentum).
Here, $\mathbf{v} = \partial_\mathbf{p}\epsilon_{\breve{\mathbf{p}}}$ is the group velocity, $e$ is the elementary charge, and $\mathbf{E}$ is the electric field created by the applied voltage $V$.
The current flowing through the channel is
\begin{equation}
I = e \int_{0}^{W}dy \int d\breve{\mathbf{p}}\,v_{x} f,
\label{eq:current}
\end{equation}
where $\int d\breve{\mathbf{p}}$ denotes summation over all states (quantum variables), and $\int_{0}^{W}dy$ is the integration across the width of the channel.
The collision integral $\mathcal{I}[f]$ has a contribution from every source of scattering and each term will be considered separately later.

The differential resistance of a two-dimensional channel can be expressed in Drude-like form through effective parameters as
\begin{equation}
\frac{dV}{dI}= \frac{L}{W} \,\rho_{xx},
\quad \frac{1}{\rho_{xx}}=e^{2}v_\mathrm{eff}\nu_\mathrm{eff} l_\mathrm{eff},
\label{eq:resistance}
\end{equation}
where $\rho_{xx}$ is the resistivity, $v_\mathrm{eff}$ is the effective quasiparticles velocity, $\nu_\mathrm{eff}$ the effective density of states, $l_\mathrm{eff}$ the effective mean free path, $W$ and $L$ are width and length of the electronic system, correspondingly.
The effective mean free path $l_\mathrm{eff}$ captures the various scattering processes that may occur in our model. 
Henceforth, we say effective and use the index ``eff'' implying that the values are calculated for a particular electron distribution.
For example, for the Fermi distribution, the effective density of states $\nu_\mathrm{eff}$ can be expressed through the actual energy-dependent density of states $\nu(\epsilon)$ as $\nu_\mathrm{eff}\approx \nu(\mathrm{max}(\mu,T))$. 
This chapter is meant to describe physical pictures of the possible regimes that can be realized in our setup.

\subsection{Scattering sources}

\label{sec:sc}

The electron transport in the channel at low temperatures is subjected to three major sources of scattering, namely a) boundary, b) impurity, and c) electron mutual scattering, each of which can be partially characterized by its mean free path.
However, these paths cannot be compared directly, and an accurate analysis is required.
In our paper, we leave the electron-phonon scattering out of scope considering the limit of low temperatures.
This limit concerns only the lattice temperature governed by the environment.
The electron gas temperature, which emerges in the electron-electron scattering rate, can be changed independently, for example when electrons are heated by a current as was done in the experiments~\cite{Molenkamp1994a,Jong1995}.

\paragraph{Boundary scattering} 

\label{sec:scwalls}

\begin{figure}
\centering
\includegraphics[width=.7\columnwidth,page=1]{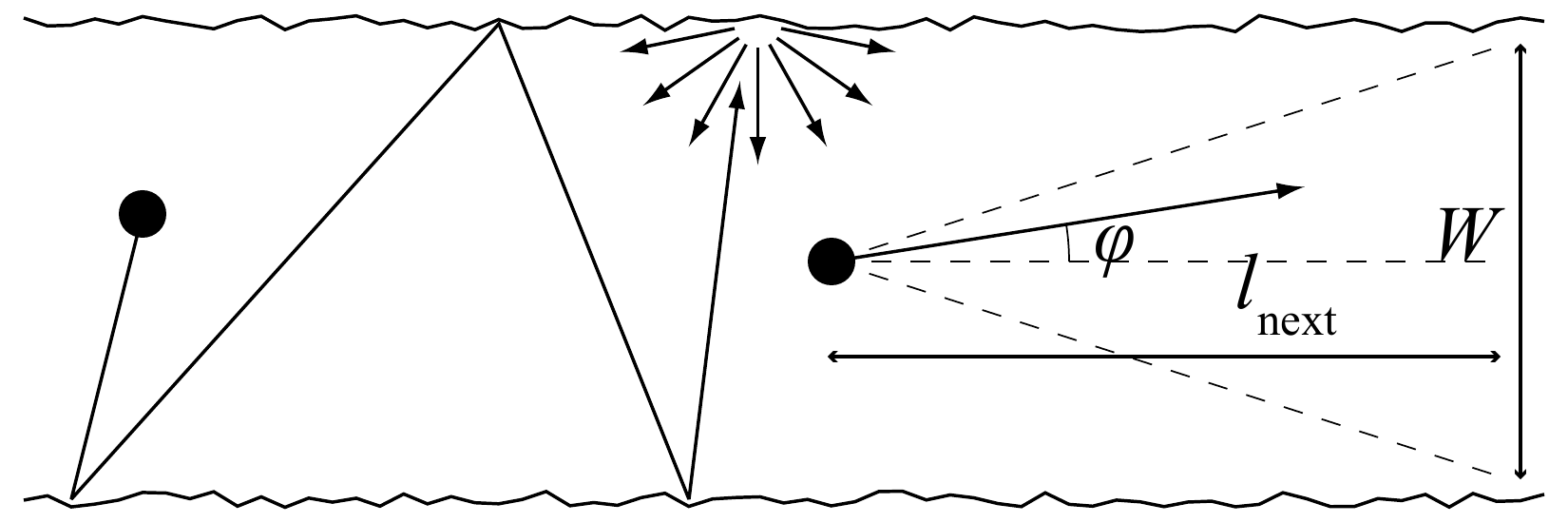}
\caption{Knudsen regime.
The schematic trajectory of the particle (black circle) hitting the ``walls'', which scatter isotropically (broken lines), is shown.
The intuitively assumed effective mean free path $l_\mathrm{eff}=W$ is affected by the small number (proportional to the angle $\varphi$) of particles propagating along the channel, which have much larger mean free path $l_{next}$.}
\label{fig:scwalls}
\end{figure}

The most natural sources of scattering are the ``walls'', i.e.\ the edges of the channel.
Here, we assume that the boundaries reflect all incoming excitations.
This requirement is plausible for true 2D materials, like monolayer graphene, for example, but should be better justified for a 3D TI with surface Dirac states.

2D Dirac excitations can, in principle, be confined in a finite area, by opening a gap outside this area.
This type of confinement, however, is only a particular case of a more general formulation of boundary conditions, which can be chirality-asymmetric and lead to the formation of unusual edge states~\cite{Ostaay2011,Kharitonov2017}.
Requiring a zero current density component perpendicular to the straight edge, the boundary condition leads to a full mirror reflection of the incident wavefunction (see Appendix~\ref{apx:dconf}).
In case of a rough edge, the incident angle has a random value at each point, reflecting the incident particle in an arbitrary direction, but with the same absolute value of the momentum.
Within the Fuchs-Sondheimer model, this implies scattering with zero specular probability~\cite{Fuchs1938,*Sondheimer1952}.
Theoretically, the Dirac states in a 3D TI emerge on the whole surface of the insulator, which has the topology of a sphere in a slab geometry.
Therefore, no boundary conditions should exist.
In experiments, however, distinct Dirac states can only be seen on the top and bottom surfaces of a thin 3D TI slab, because its edges are rather disordered.
Therefore, we believe that the coupling between top and bottom Dirac surfaces (via the side walls) can be neglected.

Hence, the incident particles are scattered isotropically at the side boundaries.
Since we conjecture that elastic disorder on the edges dominates, we come to the same conclusions as for the general Dirac equation boundary condition at the rough edge.
Thus, in our investigation we assume that the boundary scattering randomizes the incident particle momentum preserving its absolute value.

The natural (and the only) characterizing parameter of the boundary scattering process is the width of the channel $W$. 
In the case, when all other scattering sources are negligible, the effective mean free path in a channel will be
\begin{equation}
l_\mathrm{eff}\approx W\log (l_\mathrm{next}/W),
\label{eq:leffK}
\end{equation}
where $l_{next}$ is the next largest mean free path generated in our system.
The logarithm originates from the small percentage of particles that move along the channel.
If all other scattering processes generate an effective mean free path longer than the channel length $L$, then the value $l_{next}=L$ should be substituted.
In the sequel, we assume that the sample is not fully ballistic, i.e.\ there is always a finite $l_{next}$ shorter than $L$.
This regime is called the Knudsen regime~\cite{Knudsen1910} by analogy to particle diffusion in a porous medium and schematically visualized in Fig.~\ref{fig:scwalls}.

The ``walls'' cannot contribute to the collision integral directly, since they do not alter the bulk, but they affect the boundary conditions for the distribution function.
The diffusive scattering on a wall described above can be quantitatively formulated in terms of the distribution function $f(x,y, \breve{\mathbf{p}})$ at point $(x,y)$ over the states $\breve{\mathbf{p}}$ as follows:
the electrons in point $x$ at the wall moving away from this point are a) distributed isotropically and b) of the same number as those moving to this point.
Mathematically, these conditions at the bottom ($y=0$) and top ($y=W$) walls are
\begin{subequations}
\begin{align}
\left.f(x,0,\breve{\mathbf{p}})\right|_{v_{y}<0} &= \Bigl<f(x,0,\breve{\mathbf{p}}')\Bigr>_{v_{y}'>0},
\label{eq:boundary0}
\\
\left.f(x,W,\breve{\mathbf{p}})\right|_{v_{y}>0} &= \Bigl<f(x,W,\breve{\mathbf{p}}')\Bigr>_{v_{y}'<0},
\label{eq:boundaryW}
\end{align}
correspondingly.
Here, the brackets imply $\langle f \rangle_{\ldots}=S_{\ldots}[f] / S_{\ldots}[1]$, the partial averaging over the direction of the momentum, namely 
\begin{equation}
S_{v_{y}'\gtrless 0}[f] = \int \delta\bigl(\epsilon(\breve{\mathbf{p}})-\epsilon(\breve{\mathbf{p}}')\bigr) \theta\bigl( \pm v_{y}'\bigr) f d\breve{\mathbf{p}}',
\label{eq:boundaryS}
\end{equation}
\label{eq:boundary}%
\end{subequations}
where $v_{y}$ is the $y$ component of the group velocity defined above, after Eq.~\eqref{eq:boltzmann}, and $\theta$ is a Heaviside function.

\paragraph{Impurity scattering}

\label{sec:scimps}

\begin{figure}
\centering
\includegraphics[width=.7\columnwidth,page=2]{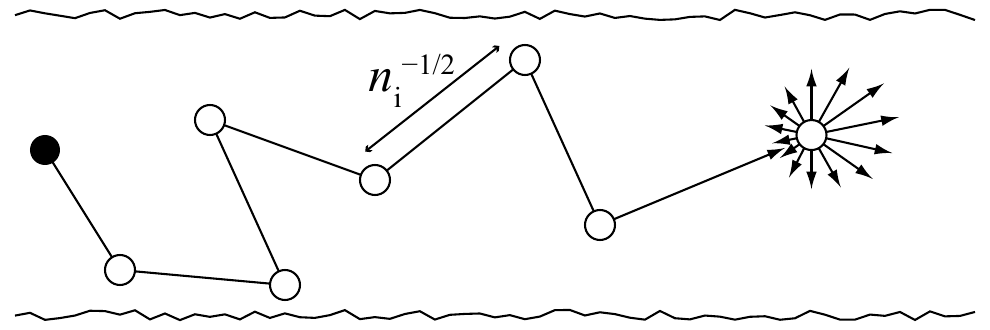}
\caption{Diffusive regime.
The particle (black circle) propagates diffusively, scattering on the impurities (white circles) the density of which is $n_\mathrm{i}$.
The non\-isotropic scattering, specific for the Dirac materials (see Sec.~\ref{sec:dirac}), is illustrated at the final scattering process.}
\label{fig:scimps}
\end{figure}

The scattering of the electrons can also take place in the whole volume of the sample due to imperfections of the material.
To simplify this analysis, we restrict ourselves to randomly distributed elastic short-range potential impurities.
The scattering rate of such impurities for a single particle with fixed energy $\epsilon$ can be estimated in Born approximation as 
\begin{equation}
\tau_\mathrm{i}^{-1}(\epsilon) \approx n_\mathrm{i}|U_{0}|^{2} \nu(\epsilon) / \hbar,
\label{eq:imprate}
\end{equation}
where $n_\mathrm{i}$ is the density of the impurities, $U_{0} = \int U(\mathbf{r})d^{2}\mathbf{r}$ with $U(\mathbf{r})$ a potential of a single impurity, and $\nu(\epsilon)$ is the density of states at energy $\epsilon$.
Thus, the effective mean free path generated by the impurities is
\begin{equation}
l_\mathrm{eff} = l_\mathrm{i} \approx \frac{\hbar v_\mathrm{eff}}{n_\mathrm{i}|U_{0}|^{2} \nu_\mathrm{eff}}.
\label{eq:leffD}
\end{equation}
where $l_\mathrm{i}$ denotes an impurity mean free path.
This regime describes standard diffusion and is schematically visualized in Fig.~\ref{fig:scimps}.

Our analysis is concentrated on short-range impurity scattering, where Eq.~\eqref{eq:imprate} is valid.
For simplicity, we consider weak scattering on local impurities within the Born approximation.
We assume a short distance $n_\mathrm{i}^{-1/2}$ between impurities compared to the system size (but large in comparison to $\lambda_\mathrm{F}$ in order to keep the Born approximation and Boltzmann kinetic equation approach valid).
This assumption implies that the graphical representation of the particle trajectory in Fig.~\ref{fig:scimps} is a rather schematic illustration, since the particle scatters not on every impurity it passes by, but rather on a small percentage of them, proportional to the scattering cross section.
It is governed by the impurity strength $U_{0}$, leading to the mean free path expression in Eq.~\eqref{eq:leffD}, which is different compared to the average distance between impurities.

The collision integral for the random impurities can be written in Born approximation as 
\begin{equation}
\mathcal{I}_\mathrm{i}[f] = \frac{\langle f \rangle - f}{\tau_\mathrm{i}},
\label{eq:Ii}
\end{equation}
where $\tau_\mathrm{i}$ is defined in Eq.~\eqref{eq:imprate} and $\langle f \rangle$ depends on the energy of the particle $\epsilon_{\breve{\mathbf{p}}}$ only, while $f$ is taken for the particular state $\breve{\mathbf{p}}$.
Angle brackets denote averaging over the direction of the momentum, defined in Eqs.~\eqref{eq:boundary0}--\eqref{eq:boundaryS} except for the missing $\theta$-function in the integrand in Eq.~\eqref{eq:boundaryS}.
In a proper treatment of Dirac materials, the scattering integral requires new terms (see below), but in the limit of the kinetic equation approach they only change the final results quantitatively~\cite{Mirlin2006,Mirlin2007}.

\paragraph{Electron-electron scattering}

\label{sec:scee}

\begin{figure}
\centering
\includegraphics[width=.7\columnwidth,page=3]{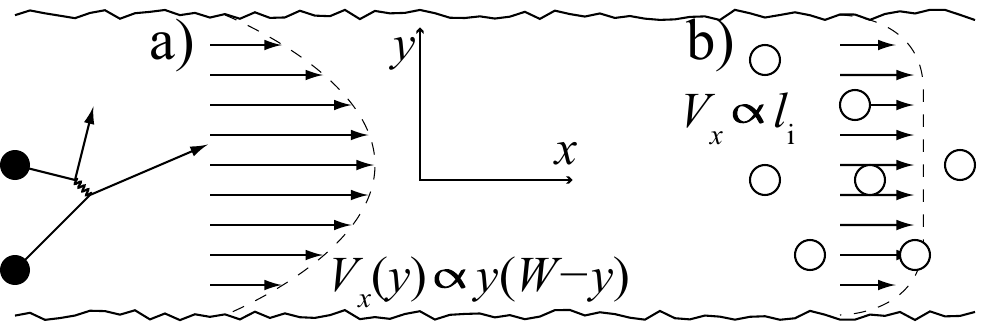}
\caption{Poiseuille regime.
Due to dominating e-e scattering, the electron system behaves as a liquid, obeying the Navier-Stokes equation.
a) The parabolic profile of the drift velocity is shown.
b) Dissipative hydrodynamic flow.
The case of weak scatterers where hydrodynamic flow is affected by the dissipative term in the Navier-Stokes equation.}
\label{fig:scee}
\end{figure}

The third, and the most sophisticated process is electron-electron (e-e) scattering.
Unlike the two others, this is an inelastic process, which redistributes momenta and energies of the colliding electrons, leading to the thermalization of the system, but, nevertheless, preserving the total momentum and energy of all electrons.
The strength of the e-e scattering can be described by the rate $\tau_\mathrm{ee}^{-1}$, or by the mean free path $l_\mathrm{ee}=v_\mathrm{eff}\tau_\mathrm{ee}$, which we will use further for the characterization of the scattering strength.

The collision integral for e-e scattering has a complicated non\-linear functional dependence on the distribution function $f$, but in certain cases it can be simplified (see Appendix~\ref{apx:eethem}).
In the collision integral  three major net values are preserved: energy $\epsilon_{\breve{\mathbf{p}}}$, momentum $\mathbf{p}$, and number of particles.
The e-e scattering tends to thermalize the system driving it to the Fermi distribution, but it does it with a twist.
The net effect is a shifted distribution function, i.e.
\begin{equation}
f(\breve{\mathbf{p}}) \stackrel{\mathcal{I}_\mathrm{ee}}{\,\,\longrightarrow\,} f_\mathrm{V}\equiv f_\mathrm{F}(\epsilon_{\breve{\mathbf{p}}} - \mathbf{V}\cdot\mathbf{p})
\label{eq:driftv}
\end{equation}
where $f_\mathrm{F}(\epsilon) \equiv 1/[e^{(\epsilon-\mu)/T}+1]$ is the Fermi distribution function (from now on, we put the Boltzmann constant to unity $k_{B}=1$).
The parameters in the distribution function---the average drift velocity $\mathbf{V}$ of the particles, the temperature $T$ of the electrons, and the chemical potential $\mu$, corresponding to the conserved quantities---must be calculated self-consistently from the initial distribution function.
The conservation laws for energy, momentum, and number of particles which allow us to calculate $T$, $\mathbf{V}$, and $\mu$, correspondingly, are
\begin{equation}
\!\!\int\!\! \epsilon_{\breve{\mathbf{p}}} \bigl[ f- f_\mathrm{V}\bigr] d \breve{\mathbf{p}} =
\!\!\int\!\! \mathbf{p} \bigl[ f- f_\mathrm{V} \bigr] d \breve{\mathbf{p}}  =
\!\!\int\!\! \bigl[ f- f_\mathrm{V}\bigr] d \breve{\mathbf{p}} = 0,
\label{eq:consee}
\end{equation}
where $f$ is the initial and $f_\mathrm{V}$ the thermalized distribution function from Eq.~\eqref{eq:driftv}.
If the system is almost thermalized, i.e.\ the difference $f- f_\mathrm{V}$ is small, one can expand the collision integral using the Callaway ansatz~\cite{Callaway1959,Gennaro1984a,*Gennaro1985a,Jong1995}:
\begin{equation}
\mathcal{I}_\mathrm{ee}[f] = \frac{f_\mathrm{V} - f}{\tau_\mathrm{ee}},
\label{eq:Iee}
\end{equation}
where the scattering time $\tau_\mathrm{ee}$ does not depend on the distribution function $f$, while the dependence on the state $\breve{\mathbf{p}}$ is very weak.
A more realistic treatment of the relaxation of the distribution function can be complicated (which we further address in Section~\ref{sec:regimes} and Appendix~\ref{apx:collin}).

At dominating e-e scattering, in comparison to other sources, the electron gas starts behaving as a fluid.
Due to the net preserving momentum, the average drift velocity $\mathbf{V}$ of the particles cannot be dissipated by the e-e collisions.
Hence, it can be treated as a flow velocity in a viscous liquid with the kinematic viscosity $\zeta \sim v_\mathrm{eff} l_\mathrm{ee}$.
Obviously, the electron-electron scattering, preserving the total momentum, cannot result in a finite resistivity on its own, so the boundary conditions play an essential role here.
Resolving the scattering on the walls within the thin layer of thickness $l_\mathrm{ee}$ near the edge, we can derive the effective conditions on larger scales: $\left.\mathbf{V}\right|_\text{boundary}=0$, the classical condition for the hydrodynamic equation.
Assuming a laminar flow, the Navier-Stokes equation yields the standard parabolic velocity profile $V_{y}=0$ and $V_{x}\propto y(W-y)$, where $x$ and $y$ are coordinates along and across the channel, correspondingly (see Fig.~\ref{fig:scee}).
The effective mean free path that will appear in the resistance in Eq.~\eqref{eq:resistance} in this case is (see the solution of Navier-Stokes equation in Appendix~\ref{apx:didro}, and its derivation in Ref.~\onlinecite{Narozhny2015})
\begin{equation}
l_\mathrm{eff} \approx \frac{W^{2}}{l_\mathrm{ee}}.
\label{eq:leffP}
\end{equation}
This case of the laminar hydrodynamic flow is called Poiseuille regime named by a scientist who investigated the cardiovascular system of a frog~\cite{Poiseuille1838}.



The described parabolic profile of the flow holds in the absence of the dissipation caused by impurities.
In the case $l_\mathrm{i}<l_\mathrm{eff}$ [where $l_\mathrm{eff}$ is predicted in Eq.~\eqref{eq:leffP}], the dissipative term lowers the flow velocity to $V_{x}\propto l_\mathrm{i}$, flattening the velocity profile 
as shown at Fig.~\ref{fig:scee}b).
Thus, the Poiseuille regime is restricted by the bound $l_\mathrm{ee}l_\mathrm{i}=W^{2}$, beyond which the mean free path is $l_\mathrm{eff}=l_\mathrm{i}$, i.e.\ the same as in the diffusive regime.
This case, however, is physically very different.
The hydrodynamical limit may still be valid, but the presence of impurities results in a flow dissipation.

%
%

\subsection{Transport regime diagram}
\label{sec:regimes}

The geometry of the sample is typically fixed in a given experiment, but the lengths $l_\mathrm{ee}$ and $l_\mathrm{i}$ can be varied by adjusting the chemical potential and temperature of the electron gas.
The chemical potential $\mu$ can be changed by applying a voltage to back or top gates.
The electron temperature $T$ can be raised by increasing the current through the sample.
This technique allows us to increase the temperature of the electron gas, leaving the bath temperature intact, hence, suppressing parasitic effects, such as electron-phonon scattering.
We develop a ``phase diagram'' that will allow us to classify the possible transport regimes.
If we could scale the e-e scattering and impurity mean free paths independently, say, $l_\mathrm{ee}$ by one external parameter and $l_\mathrm{i}$ by another one, then a simplified diagram would look like Fig.~\ref{fig:roadmap}a).
Two main lines in the figure correspond to the crossovers \circled{1} $l_\mathrm{i}=W$ and \circled{2} $l_\mathrm{ee}=W$.
The third line indicates when the e-e and impurity scattering are equal, \circled{3} $l_\mathrm{ee}=l_\mathrm{i}$.

\subsubsection{Dirac spectrum}
\label{sec:dirac}

Dirac materials are characterized by a distinct Hamiltonian $H=v \,\boldsymbol{\sigma}\cdot\mathbf{p}$, where $\boldsymbol{\sigma}$ are  Pauli matrices acting on the space of (iso-)spin and $v$ is the Dirac velocity.
The spectra of such Hamiltonians consists of two---conduction and valence---cone-shaped bands with the dispersion relation $\epsilon_{+,p}=vp$ and $\epsilon_{-,p}=-vp$, correspondingly.
The density of states in 2D Dirac materials is linear with energy $\nu(\epsilon) = |\epsilon|/2\pi\hbar^{2}v^{2}$.
The effective velocity in Dirac media is the constant Dirac velocity $v_\mathrm{eff}=v$.
The effective density of states and the e-e scattering rate depend on the distribution function $f_{\pm,\mathbf{p}}$, which is derived from the standard kinetic equation below~\cite{Fritz2008,Kechedzhi2008,Kashuba2008}.
We assume that the quasiparticles are thermalized and have a distribution function close to a Fermi distribution:
\begin{equation}
f_{\pm,\mathbf{p}} \approx f_\mathrm{F}(\epsilon_{\pm,p}) \equiv \frac{1}{e^{(\epsilon_{\pm,p}-\mu)/T}+1}.
\end{equation}
In the two opposite cases of strong ($\mu\gg T$) and weak ($\mu\ll T$) chemical potential the effective energies of an electron are $\mu$ and $T$, respectively.

\begin{figure}
\centering
\includegraphics[width=\columnwidth]{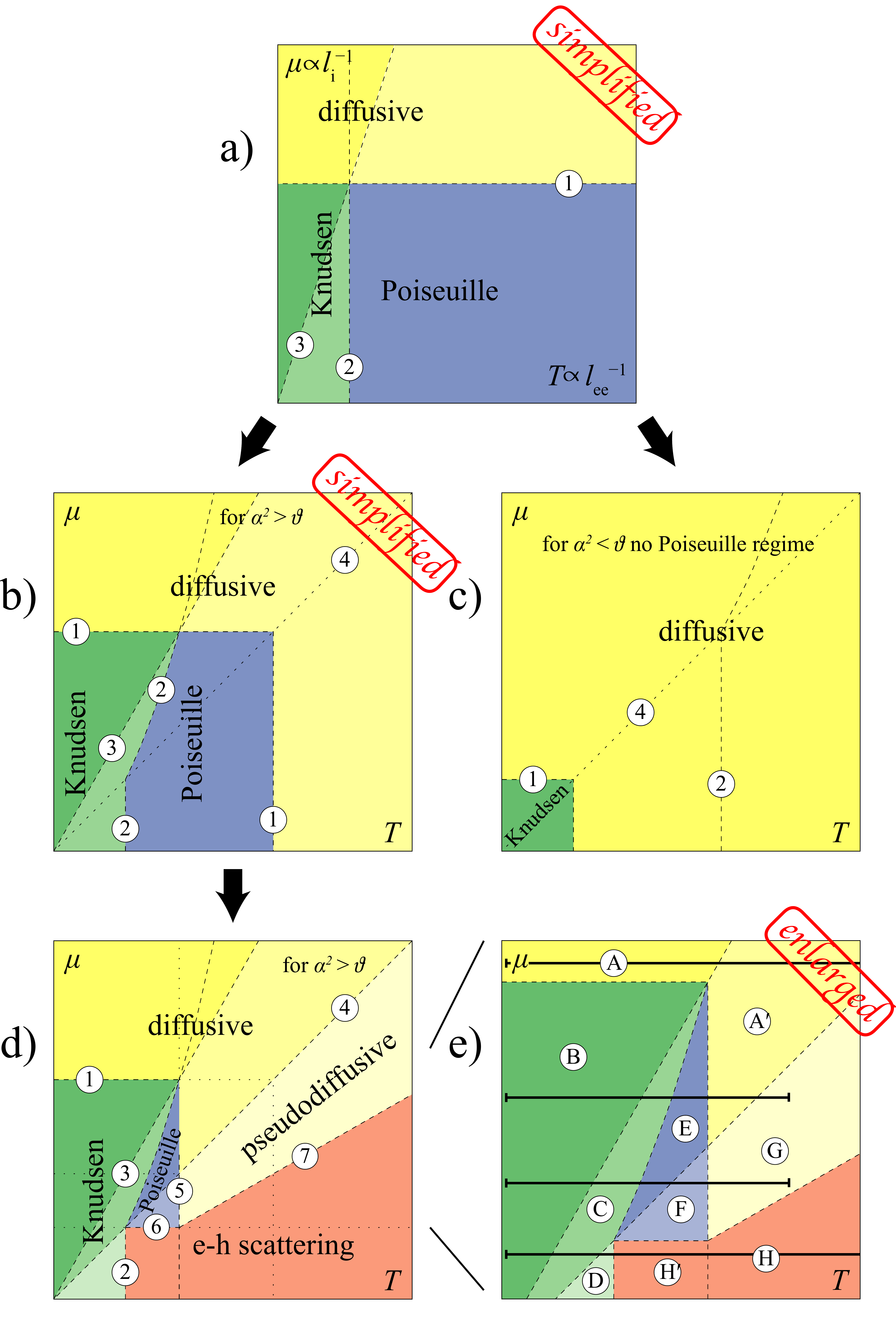}
\caption{Set of ``phase diagrams'' helping us to introduce the different transport regimes in Dirac channels.
a) The simplest diagram introduces three basic transport regimes depending on ratios between e-e scattering ($l_\mathrm{ee}$), impurity scattering ($l_\mathrm{i}$), and the width of the channel $W$.
For pedagogical reasons, first simplified dependencies of $l_{i/ee}$ on $\mu$ and $T$ are assumed [see the axes labels].
b) Improvement of the previous diagram by taking into account more realistic physical dependencies of $l_{i/ee}$ on $\mu$ and $T$ for clean ($\vartheta < \alpha^{2}$) samples.
The diagrams a) and b) are labeled ``simplified'' since they are based on simplified assumptions that allow us to disentangle different scattering processes.
c) Physical diagram for dirty ($\vartheta > \alpha^{2}$) samples.
d) Physical diagram for clean ($\vartheta < \alpha^{2}$) samples derived from case b) by taking into account electron-hole scattering and the sensitivity of hydrodynamical flow on impurities.
The Knudsen regime is labeled by green color, Poiseuille regime by blue, diffusive regime by yellow, and the area with dominating electron-hole scattering by red.
The different shades of the color denote slight differences in the conductance dependence on $\mu$ and $T$ within the same regime.
For instance, light yellow stands for the (intermediate) pseudo\-diffusive regime.
The formulas for the lines labeled in d) are collected in Table~\ref{tab:crossovers}.
The differential resistances for each regime labeled in e) are collected in Table~\ref{tab:regimes}.
The intervals marked by the thick black lines in e) correspond to the parameter intervals for the plots in Fig.~\ref{fig:cross}.
}
\label{fig:roadmap}
\end{figure}

This implies that the dependence on $\mu$ and $T$ of the effective density of states is 
\begin{equation}
\nu_\mathrm{eff} = \frac{1}{2\pi\hbar^{2}v^{2}}\times 
\begin{cases}
|\mu| & \text{for $\mu\gg T$,}\\
T & \text{for $\mu\ll T$.}
\end{cases}
\label{eq:nueff}
\end{equation}
Therefore, in our model, the impurity mean free path can be described by the formula
\begin{equation}
l_\mathrm{i} = \vartheta^{-1} \frac{\hbar v}{\mathrm{max}(\mu, T)},
\label{eq:li}
\end{equation}
which is valid for the two regimes $\mu\gg T$ and $\mu\ll T$.
Here, the dimensionless parameter $\vartheta \sim n_\mathrm{i} |U_{0}|^{2}/\hbar^{2}v^{2}$ characterizes the cleanliness of the sample.
The modified dependence of the impurity scattering on $\mu$ and $T$ is reflected in Figs.~\ref{fig:roadmap}b)~and~c) by the bending of the line \circled{1} over the line \circled{4} that separates the $\mu\gg T$ and $\mu\ll T$ cases.
The (iso-)spin-momentum coupling in the Dirac Hamiltonian prohibits backscattering.
In 1D materials, this leads to non\-dissipative propagation, but, in the 2D case, particles can scatter at all angles except for 180$^{\circ}$, schematically illustrated on the last impurity in Fig.~\ref{fig:scimps}.
Thus, the scattering integral for electrons with energy $\epsilon_{\pm,\mathbf{p}}$ gets modified in comparison to Eq.~\eqref{eq:Ii} as~\cite{Kechedzhi2008}
\begin{equation}
\mathcal{I}_\mathrm{i}[f]=\frac{\langle f\rangle + \mathbf{n}\langle \mathbf{n}' f\rangle - f}{\tau_\mathrm{i}}
\label{eq:Ii2dd}
\end{equation}
where $\mathbf{n} = \mathbf{v}/v = \pm\mathbf{p}/p$, and the quantities $\langle f\rangle$, $\langle \mathbf{n}' f\rangle$, and $\tau_\mathrm{i}(\epsilon) = 1/\vartheta|\epsilon|$ depends on the energy $\epsilon_{\pm,\mathbf{p}}$ only (see Appendix~\ref{apx:limax}).


The electron-electron scattering does also crucially depend on the chemical potential of the Dirac material.
In the regime $\mu\gg T$, the system mimics a 2D electron gas and the e-e scattering mean free path is~\cite{Gennaro1984a,*Gennaro1985a,Fritz2008,Muller2008b}
\begin{equation}
l_\mathrm{ee} \sim \alpha^{-2} \hbar v \frac{\mu}{T^{2}}\qquad\text{(for $\mu\gg T$).}
\label{eq:leemu}
\end{equation}
A particularly interesting behavior of e-e scattering can be observed for $\mu\ll T$:
it reminds us of the thermalization dynamics in bad metals~\cite{Fritz2008,Siegel2013}, i.e.\
\begin{equation}
l_\mathrm{ee} \sim \alpha^{-2} \frac{\hbar v}{T} \qquad\text{(for $\mu\ll T$).}
\label{eq:leeT}
\end{equation}
This regime is reflected by the modified behavior of the $l_\mathrm{ee}=W$ line labeled as \circled{2} in Fig.~\ref{fig:roadmap}b)~and~c).
Here, $\alpha$ is the parameter of the electromagnetic field coupling.
In the absence of screening and other renormalization effects, it would be equal to the effective fine structure constant $\alpha_{0} \sim e^{2}/\hbar v$, which due to the low Dirac velocity compared to the speed of light $c/v\sim 300$ can in principle be large ($\alpha_{0}\sim 3$) in condensed matter systems.
In reality, the coupling constant is however strongly renormalized by (dumped) plasmon screening~\cite{Foster2008}, in the RG sense~\cite{Sheehy2007}, by dielectric properties of the substrate, etc.~\cite{Kotov2012}
In any case, $\alpha$ is typically neither an extremely large nor small number in known Dirac materials.

The logarithmic enhancement of collinear electron-electron scattering in Dirac materials due to the linear spectrum~\cite{Briskot2015} can be estimated as $\tau_{cee}^{-1}\sim\tau_{ee}^{-1}\log \alpha^{-2}$ for small $\alpha$.
In our investigation, we do not consider this particular type of scattering for two reasons.
At first, if $\alpha\sim1$, which seems to be experimentally relevant, the logarithm $\log\alpha$ does not lead to any substantial enhancement in e-e scattering. 
Second, the collinear e-e scattering leads to the relaxation of the energy (and correspondingly the absolute value of the momentum), but not the relaxation of the momentum direction, which plays the dominant role in transport.
We address this issue in details in the Appendix~\ref{apx:collin}.

%
%

Note, that the dependence of the e-e and impurity scattering on $\mu$ and $T$ implies the existence of the line \circled{3}, where $l_\mathrm{ee}=l_\mathrm{i}$, only for clean enough samples, more precisely, if $\vartheta < \alpha^{2}$, see Fig.~\ref{fig:roadmap}b).
Otherwise, if $\vartheta > \alpha^{2}$, impurity scattering always dominates, i.e.\ $l_\mathrm{i}<l_\mathrm{ee}$ for all values of $\mu$ and $T$, see Fig.~\ref{fig:roadmap}c).


The diagram b) in Fig.~\ref{fig:roadmap} does not take into account the flow dissipation due to the impurities discussed at the end of Section~\ref{sec:sc}.
The corresponding boundary of the Poiseuille regime is given by the relation $l_{i}l_{ee}=W^{2}$ manifested by the line \circled{5} in Fig.~\ref{fig:roadmap}d).
%
%
%
For the case $\mu\gg T$, as follows from Eqs.~\eqref{eq:li} and~\eqref{eq:leemu}, the condition $l_\mathrm{ee}l_\mathrm{i}=W^{2}$ transforms into the relation $T=\alpha\sqrt\vartheta\,\hbar v/W$, represented by line \circled{5} in Fig.~\ref{fig:roadmap}d).
In the opposite case $T\gg\mu$, the condition $l_\mathrm{ee}l_\mathrm{i}=W^{2}$ results into the same relation for temperature (up to a numerical factor).
However, the validity of this condition itself has to be justified because of the significant presence of hole\-like excitations, which substantially changes the physical picture.


\begin{table}[h]
\centering
\def\arraystretch{2.25}
\begin{tabular*}{\linewidth}{l@{\hspace{2em}}l@{\hspace{2em}}@{\hspace{2em}}l@{\extracolsep{\fill}}}
\hline\hline
\circled{1} & $l_\mathrm{i}=W$           & $\displaystyle \mu=\frac{1}{\vartheta}\frac{\hbar v}{W}$ \\[5pt]
\circled{2} & $l_\mathrm{ee}=W$          & $\hspace{-1em}\displaystyle 
\begin{cases}
\displaystyle T=\frac{1}{\alpha^{2}}\frac{\hbar v}{W} & \text{if $\mu\ll T$}
\\[5pt]
\displaystyle \mu=\alpha^{2}T^{2}\frac{W}{\hbar v} & \text{if $\mu\gg T$}
\end{cases}
$ \\
\circled{3} & $l_\mathrm{ee}=l_\mathrm{i}$      & $\displaystyle \mu=\frac{\alpha}{\sqrt\vartheta}T$ \\[5pt]
\hline
\circled{4} & & $\mu=T$\\[4pt]  
\hline
\circled{5} & $l_\mathrm{ee} l_\mathrm{i} =W^{2}$ & $\displaystyle T=\frac{1}{\alpha\sqrt\vartheta}\frac{\hbar v}{W}$ \\[5pt]
\hline
\circled{6} & $\displaystyle \frac{l_\mathrm{ee}}{W} = \frac{\mu}{T}$  & $\displaystyle \mu=\frac{1}{\alpha^{2}}\frac{\hbar v}{W}$ \\
\circled{7} & $\displaystyle \frac{l_\mathrm{ee}}{l_\mathrm{i}} = \frac{\mu^{2}}{T^{2}}$ & $\displaystyle \mu=\frac{\sqrt\vartheta}{\alpha}T$ \\[5pt]
\hline\hline
\end{tabular*}
\caption{Crossovers labeled by encircled numbers in Fig.~\ref{fig:roadmap}d).
The first block classifies the cases of the lengths $l_\mathrm{i}$, $l_\mathrm{ee}$, and $W$ ordering;
the second block distinguishes between $\mu\ll T$ and $\mu\gg T$ cases;
the third one corresponds to the limit of the Poiseuille regime;
and the fourth one describes the effect of the electron-hole scattering in the regime $\mu\ll T$.}
\label{tab:crossovers}
\end{table}
\begin{table}[h]
\centering
\def\arraystretch{2.25}
\begin{tabular*}{\linewidth}{l@{\hspace{5em}}r@{\hspace{2pt}}l@{\extracolsep{\fill}}}
\hline\hline
\circled{A},\circled{A\makebox[0pt]{$^\prime$}} & $\displaystyle \rho_{xx}= \frac{h}{e^{2}} \times$ & $\displaystyle \vartheta$							\\[5pt]
\hline
\circled{B} & $\times$ & $\displaystyle\frac{\hbar v}{\mu W}\log^{-1}\frac{\hbar v}{\vartheta\mu W}$					\\
\circled{C} & $\times$ & $\displaystyle\frac{\hbar v}{\mu W}\log^{-1}\frac{\mu \hbar v}{\alpha^{2} T^{2} W}$	\\
\circled{D} & $\times$ & $\displaystyle\frac{\hbar v}{ T W }\log^{-1}\frac{\hbar v}{\alpha^{2} T W}$			\\[5pt]
\hline
\circled{E} & $\times$ & $\displaystyle \alpha^{-2} \frac{\hbar^{2} v^{2}}{T^{2} W}$							\\
\circled{F} & $\times$ & $\displaystyle \alpha^{-2} \frac{\hbar^{2} v^{2}}{\mu^{2} W}$							\\[5pt]
\hline	
\circled{G} & $\times$ & $\displaystyle \vartheta\frac{T^{2}}{\mu^{2}}$												\\
\circled{H},\circled{H\makebox[0pt]{$^\prime$}} & $\times$ & $\displaystyle \alpha^{2}$																\\[5pt]
\hline\hline
\end{tabular*}
\caption{The resistivity for the regimes labeled by encircled letters in Fig.~\ref{fig:roadmap}e).
The blocks are ordered according to the described regimes: (from top to bottom) diffusive, Knudsen, Poiseuille, and dominating e-h scattering (including pseudo\-diffusive) regimes.}
\label{tab:regimes}
\end{table}

\subsubsection{Electron-hole scattering}
\label{sec:diraclee}

The Dirac spectrum consists of conduction and valence bands occupied by electrons and holes, respectively.
The ratio of chemical potential and temperature determines whether the system has one or two types of charge carriers.
In the first case, the system behaves similar to a Fermi liquid.
In the other case, an additional electron-hole (e-h) scattering emerges, which at certain circumstances may lead to the Dirac liquid behavior~\cite{Fritz2008}.


If $T\simeq\mu$, in addition to electron excitations (in the conduction band), the system has a substantial amount of hole excitations (in the valence band), which have anticollinear momentum and velocity, opposite to the electrons.
The mutual electron-hole scattering is known to create a finite resistivity in the Dirac system in the degenerate limit ($\mu=0$)~\cite{Kashuba2008}.
If $l_\mathrm{ee}$ is the shortest length scale, the mean free path is
\begin{equation}
l_\mathrm{eff}=l_\mathrm{ee}.
\label{eq:leffeh}
\end{equation}
The ratio of e-e and e-h collision for finite $\mu$ can be estimated by the volumes in reciprocal space taken by electrons $n_{e}$ and holes $n_{h}$ as $(n_{e}-n_{h})/(n_{e}+n_{h})$ [see Appendix~\ref{apx:didro}].
In case $\mu\ll T$, the ratio will be $\mu^{2}/T^{2}$.
Thus, the system is composed of two parallel channels that contribute to the conductivity: the e-h scattering dominated channel and e-e collision driven hydrodynamic flow, which is suppressed proportionally to the rate of the e-e to e-h collisions.
In the absence of impurity scattering the mean free path is then
\begin{equation}
l_\mathrm{eff}=l_\mathrm{ee}+\frac{W^{2}}{l_\mathrm{ee}}\times \frac{\mu^{2}}{T^{2}}.
\end{equation}
The line \circled{7}, where these terms are comparable is given by the relation $l_\mathrm{ee}=W \mu /T$ and according to Eq.~\eqref{eq:leeT} sets the critical chemical potential $\mu=\alpha^{-2}\hbar v /W$, see Fig.~\ref{fig:roadmap}d).
Below this line is a e-h scattering dominated regime with the mean free path given in Eq.~\eqref{eq:leffeh}.
Between line \circled{7} and line \circled{4}, which denotes the crossover to the regime $\mu\gg T$, we find a pseudo-Poiseuille regime, where the hydrodynamical flow dominates and obeys the very same Navier-Stokes equation, but its contribution is suppressed by the factor $\mu^{2}/T^{2}$.
This results in the mean free path
\begin{equation}
l_\mathrm{eff}=\frac{W^{2}}{l_\mathrm{ee}}\frac{\mu^{2}}{T^{2}}.
\label{eq:leffpP}
\end{equation}
This special regime is further restricted from the third side by the increasing strength of impurity scattering, entering in the very same way as in the Navier-Stokes equation.
Hence, we substitute $W^{2}/l_\mathrm{ee} \to l_\mathrm{i}$ if $l_\mathrm{i}<W^{2}/l_\mathrm{ee}$.
This means that the line \circled{5} that separates the Poiseuille regime from diffusive regimes preserves its definition and shape under the line \circled{4} (i.e.\ for $\mu\ll T$).

If the impurity scattering is very strong, but still $l_\mathrm{ee}<l_\mathrm{i}$, it transforms the Poiseuille hydrodynamic flow into dissipative hydrodynamic flow yielding a mean free path of
\begin{equation}
l_\mathrm{eff}=l_\mathrm{ee}+l_\mathrm{i}\times \frac{\mu^{2}}{T^{2}}.
\end{equation}
For the case $\mu\gg T$, the e-e scattering takes place in the conduction band, so the suppressing factor $\mu^{2}/T^{2}$ drops out.
Since $l_\mathrm{ee}<l_\mathrm{i}$ in the clean case, the second term always dominates for $\mu\gg T$, as illustrated in Fig.~\ref{fig:roadmap}d).
For $\mu\ll T$, it is not always the case.
Nevertheless, if $l_\mathrm{ee}/l_\mathrm{i}<\mu^{2}/T^{2}$, the second term dominates after all, giving us a line \circled{7} described by the relation $\mu = \alpha^{-1}\sqrt\vartheta \,T$, which singles out a special pseudo\-diffusive regime with effective mean free path
\begin{equation}
l_\mathrm{eff}=l_\mathrm{i} \frac{\mu^{2}}{T^{2}}.
\label{eq:leffpd}
\end{equation}
Knowing the estimates for the effective mean free path for every regime [Eqs.~\eqref{eq:leffK}, \eqref{eq:leffD}, \eqref{eq:leffP}, \eqref{eq:leffeh}, \eqref{eq:leffpP}, \eqref{eq:leffpd}], and the expression for the density of states from Eq.~\eqref{eq:nueff}, using the formula in Eq.~\eqref{eq:resistance}, we can determine the differential resistance for each case.
We have summarized the definitions of all lines shown in Fig.~\ref{fig:roadmap}d) and their expressions for the $\mu$--$T$ dependence in Table~\ref{tab:crossovers} and the differential resistances for all regimes shown in the enlarged diagram in Fig.~\ref{fig:roadmap}e) in Table~\ref{tab:regimes}.

\section{Kinetic equation for the transport in a Dirac channel}
\label{sec:kinetic}

In this section, we demonstrate the solution of the kinetic equation given in Eq.~\eqref{eq:boltzmann} with a collision integral which contains both electron-electron and impurity scattering contributions, see Eqs.~\eqref{eq:Iee} and~\eqref{eq:Ii2dd}, correspondingly, together with the boundary conditions formulated in Eqs.~\eqref{eq:boundary0}--\eqref{eq:boundaryS}. 

It is convenient to parametrize a state belonging to the Dirac spectrum by a direction of the velocity vector $\mathbf{n}$ and an energy of the state $\epsilon$.
The sign of the energy denotes the band, whilst the velocity and momentum can be expressed as $\mathbf{v}=v\mathbf{n}$ and $\mathbf{p}=\epsilon\,\mathbf{n}/v$.
Together with the explicit expression of the collision integrals from Eqs.~\eqref{eq:Iee} and~\eqref{eq:Ii2dd}, the static kinetic equation takes the form
\begin{multline}
\mathbf{v} \cdot \nabla f_{\mathbf{n},\epsilon} + e\mathbf{E} \cdot \partial_{\mathbf{p}} f_{\mathbf{n},\epsilon} = \\=
\frac{\langle f_{\mathbf{n},\epsilon} \rangle + \langle \mathbf{n}' f_{\mathbf{n}',\epsilon} \rangle \mathbf{n} - f_{\mathbf{n},\epsilon}}{\tau_\mathrm{i}(\epsilon)} 
+\\+ \frac{f_\mathrm{F}(\epsilon - \mathbf{V}\cdot\mathbf{n}\,\epsilon /v)-f_{\mathbf{n},\epsilon}}{\tau_\mathrm{ee}},
\label{eq:kineqfull}
\end{multline}
where the momentum derivative in terms of the new parametrization turns into $v^{-1}\partial_{\mathbf{p}}=\mathbf{n}\partial_{\epsilon} + \epsilon^{-1} \partial_{\mathbf{n}}$,
and the drift velocity $\mathbf{V}$ is obtained from the second condition in Eq.~\eqref{eq:consee}, which corresponds to momentum conservation.
We consider the case when the leads attached to the ends of the channel have the same electron temperature.
Thus, the energy conservation condition is satisfied by definition.
The conservation of the number of particles can be automatically satisfied using the following parametrization of the distribution function:
\begin{equation}
f = f_\mathrm{F}(\epsilon+\chi)=\frac{1}{e^{(\epsilon+\chi-\mu)/T}+1}.
\end{equation}
In the experimentally relevant case, the deviation of the distribution function from the Fermi distribution is typically weak, i.e.\ $|\partial_{\mathbf{p}}\chi| \ll v $.
Under this condition, the expansion $f_\mathrm{F}(\epsilon+\chi) \approx f_\mathrm{F}+f_\mathrm{F}'\chi$, where $f_\mathrm{F}=f_\mathrm{F}(\epsilon)$ and $f_\mathrm{F}'=\partial_{\epsilon}f_\mathrm{F}(\epsilon)$, is valid (see Appendix~\ref{apx:expansion}).
Performing this expansion in the kinetic equation, Eq.~\eqref{eq:kineqfull} becomes
\begin{multline}
\mathbf{n}\cdot \left( \nabla\chi - \nabla\mu - e \mathbf{E} \right) =
 \vartheta \frac{|\epsilon|}{\hbar v}\Bigl( \langle\chi \rangle + \langle\chi \mathbf{n}' \rangle \mathbf{n} -\chi \Bigr) 
 +\\+ l_\mathrm{ee}^{-1}\Bigl(\overline{\langle\chi \rangle} + \epsilon v^{-1} \mathbf{V}\cdot\mathbf{n} - \chi\Bigr),
\label{eq:kinchi}
\end{multline}
and the boundary conditions are readily obtained from Eq.~\eqref{eq:boundary} by substituting $f\to\chi$.
Here, the overline is defined as
\begin{equation}
\overline{X}=F[X]/F[1], \,\,\,\,\text{where}\,\,\,\, F[X]=-\int X\, f_\mathrm{F}' |\epsilon|d\epsilon.
\label{eq:defjol}
\end{equation}
Note that $F[1]=\int_{0}^{+\infty}f_\mathrm{F}d\epsilon-\int_{-\infty}^{0}(1-f_\mathrm{F})d\epsilon = \mu$.
The solution of Eq.~\eqref{eq:kinchi} can be exploited for the calculation of the resistivity through the formula for the current density
\begin{equation}
\mathbf{j}= \frac{e}{2\pi \hbar^{2} v} F[ \langle \mathbf{n} \chi \rangle ] = \frac{e}{2\pi \hbar^{2} v}\mu\, \overline{ \langle \mathbf{n} \chi \rangle }.
\label{eq:defjolj}
\end{equation}
The value $\chi$ as a function of the direction $\mathbf{n}$ does not have a zero harmonic contribution i.e.\ $\langle \chi \rangle=0$.
This stems from the kinetic equation~\eqref{eq:kinchi} being linear in $\chi$ possessing certain reflection symmetries (see Appendix~\ref{apx:symmetry}), and satisfying particle conservation by construction.
The momentum conservation law takes the form
\begin{equation}
\underline{\langle \mathbf{n} \chi \rangle } = - \frac{1}{2v}\mathbf{V},
\qquad
\underline{X} = \left.F[X\,\epsilon] \right/ F[\epsilon^{2}].
\label{eq:defVul}
\end{equation}
We assume a uniform chemical potential for simplicity, i.e.\ $\nabla\mu=0$.
The current is driven by the electric field $\mathbf{E}=(E,0)$ directed along the channel.
The linearity of Eq.~\eqref{eq:kinchi} in $\chi$ allows us to parametrize the function  by the coordinate and momentum dependent effective mean free path $\chi = e E \cos\varphi\, l(y,\epsilon,\varphi)$.
Note that the effective mean free path $l$ depends on the coordinate across the channel $y$, the energy $\epsilon$, and the angle $\varphi$ shown in Fig.~\ref{fig:scwalls}.
The conductivity, using the function $l$ can then be expressed as
\begin{equation}
\sigma \!=\! \frac{e^{2}}{2\pi \hbar} \frac{\mu}{\hbar v W} \!\! \int_{0}^{W} \!\! \overline{\tilde{l}\,}dy,
\quad
\tilde{l}(y,\epsilon) \!=\! 2 \langle l(y,\epsilon,\varphi)\cos^{2}\varphi \rangle,
\label{eq:condtl}
\end{equation}
where the bar over $\tilde{l}$ is defined in Eq.~\eqref{eq:defjol}.
Since, according to Eq.~\eqref{eq:condtl}, the full angle-resolved information about the distribution function is not needed, we restrict ourselves to the following equation for $\tilde{l}$ (see Appendix~\ref{apx:leqn}):
\begin{multline}
\tilde{l}(y,\epsilon) = 
\int_{0}^{W}\left(1+\frac{\vartheta}{2\hbar v} |\epsilon| \tilde{l}(y',\epsilon) + \frac{\epsilon}{l_\mathrm{ee}}\underline{\tilde{l}(y',\epsilon')}\right)
\times\\\times
K\bigl(|y-y'|/l_\mathrm{tot}\bigr) dy',
\label{eq:tlK}
\end{multline}
where $K(z) = \frac{2}{\pi} \int_{0}^{\pi/2} \frac{\cos^{2}\varphi}{\sin\varphi} e^{-z/\sin\varphi}d\varphi$, the parameter 
$l_\mathrm{tot}^{-1} = \frac{\vartheta}{\hbar v} |\epsilon| + l_\mathrm{ee}^{-1}$, and the function $\underline{\tilde{l}(y',\epsilon')}$ depends on $y'$ only [the underline is defined in Eq.~\eqref{eq:defVul}]. 
This Fredholm equation of a second kind can be solved numerically, for example as described in Appendix~\ref{apx:numerical}.
We discuss the results for the resistivity of the channel in the next section.

\section{Results and discussion}
\label{sec:results}

\begin{figure}
\centering
\includegraphics[width=.78\columnwidth]{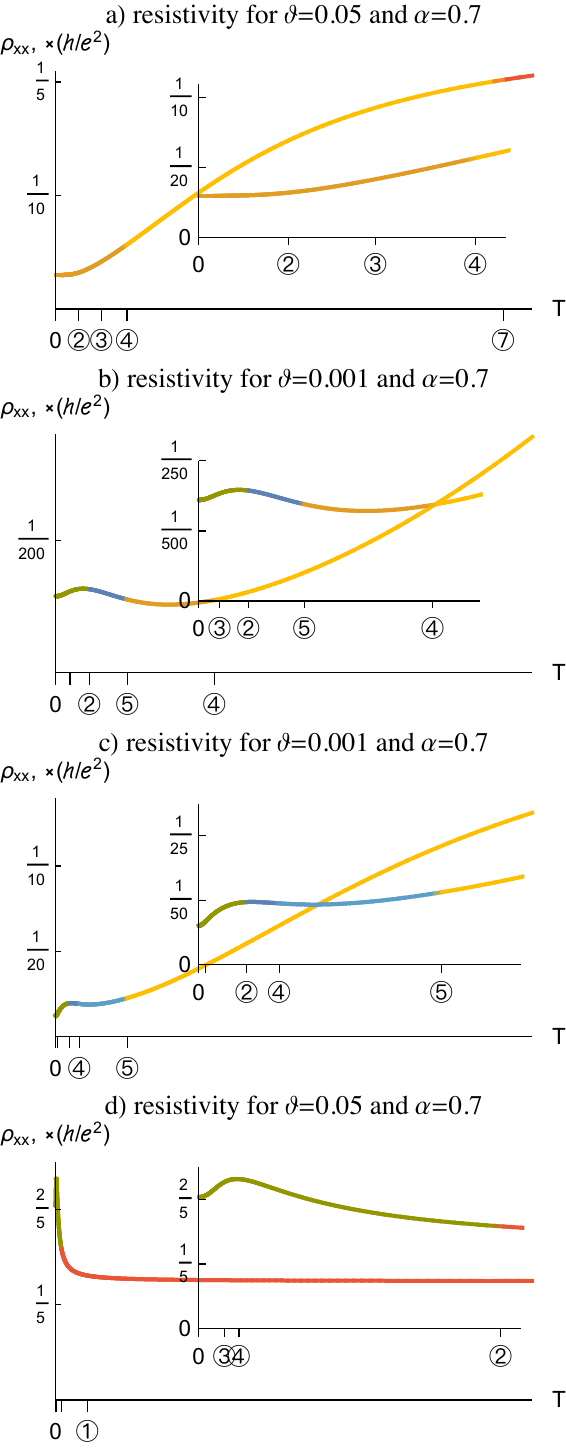}
\caption{Temperature dependence of the resistivity for different fixed chemical potentials corresponding to the thick horizontal lines in Fig.~\ref{fig:roadmap}e).
The insets contain the enlarged features of the main plot.
The circled labels on $x$ axes correspond to the lines in Fig.~\ref{fig:roadmap}d) and the color of the curve sections matches the color associated to the corresponding regime in Figs.~\ref{fig:roadmap}d)~and~e).
The labels on the $y$ axis are the values of the resistivity in units of the Klitzing resistance $h/e^{2}$.
All plots are computed for the e-e coupling $\alpha=0.7$.
Plots a) and d) are drawn for $\vartheta=0.05$, and plots b) and c) are generated for the much cleaner case $\vartheta=10^{-3}$ which resolves the Poiseuille regime (see Fig.~\ref{fig:biases} and its discussion in Section~\ref{sec:results} concerning the existence of the distinct hydrodynamic regime).
The scale of the $x$ axis of the main plots that belong to the same pair [with the same $\alpha$ and $\vartheta$, i.e.\ a,d) and b,c)] is kept constant.
}
\label{fig:cross}
\end{figure}

To illustrate the transport properties of our system we supplement our qualitative picture in Fig.~\ref{fig:roadmap} with the quantitative  solution of the kinetic equation described in Section~\ref{sec:kinetic} using a numerical computation (see Appendix~\ref{apx:numerical} for details).
For a given choice of sample cleanliness $\vartheta$ and e-e coupling $\alpha$, we calculate the resistivity as a function of temperature $T$ for different values of the chemical potential $\mu$.
The parameter range of the chemical potential and temperature used for the computation is qualitatively shown in the Fig.~\ref{fig:roadmap}e) by the thick black horizontal lines.
The plots are collected in Figs.~\ref{fig:cross}a)--d) sorted in the same order as the lines in the Fig.~\ref{fig:roadmap}e).
The different regimes in a single plot are designated by the color of the curve that corresponds to the colors used in Fig.~\ref{fig:roadmap}d)~and~e).
The most interesting features are enlarged in insets.
Let us discuss these regimes and compare the behavior of the curves with the qualitative predictions summarized in Table~\ref{tab:regimes}.

The simplest case is the one with the largest chemical potential.
It is demonstrated in Fig.~\ref{fig:cross}a) for $\vartheta=0.05$ and $\alpha=0.7$.
The chemical potential is large enough to make $l_\mathrm{i}$ shorter than the width of the channel and we see the constant value (approximately) of the resistance predicted by expression $\circled{A}$ in Table~\ref{tab:regimes}.
For temperatures larger than the chemical potential (to the right of the tick $\circled{4}$) we enter the pseudo\-diffusive regime, where the number of the effective carriers drops and the resistivity rises according to formula $\circled{G}$ in Table~\ref{tab:regimes}.
This rise takes place until the line $\circled{7}$, after which the e-h scattering dominates.
The resistance then saturates to the value given by the expression $\circled{H}$.
Naturally, the e-h scattering driven resistance $\circled{H}$ is much easier to be reached at low values of the chemical potential, e.g.\ in the case of Fig.~\ref{fig:cross}d), which is drawn for the same values of $\vartheta$ and $\alpha$, and with the same scale on the $x$-axis as the Fig.~\ref{fig:cross}a).
At low temperatures, the resistance is large due to the low density of states and depends weakly on  temperature (see $\circled{B}$ and $\circled{C}$ in Table~\ref{tab:regimes}), but after the crossing label $\circled{4}$ the density of states is proportional to the temperature (regime $\circled{D}$) and the resistivity drops until the temperature reaches the value $\circled{2}$, above which the regime $\circled{H}$ takes over.

In order to resolve the Poiseuille regime in Figs.~\ref{fig:cross}b)~and~c) we investigate a much cleaner system with $\vartheta=10^{-3}$ and the same choice of $\alpha$.
As one can see, smaller $\vartheta$ moves the lines $\circled{1}$ and $\circled{5}$ upwards and to the right, correspondingly, leaving the lines $\circled{2}$ and $\circled{6}$ intact, increasing the regions $\circled{E}$ and $\circled{F}$.

The plot in Fig.~\ref{fig:cross}b) is a prominent illustration of the relativistic Gurzhi effect in a Dirac material~\cite{Gurzhi1963,*Gurzhi1964,*Gurzhi1968}, which constitutes a non\-monotonic dependence of the resistivity on the strength of the e-e scattering regulated by the electron temperature.
At low temperature, the system is in the Knudsen regime.
Then, the particles relax their momentum by scattering at the walls, see regimes $\circled{B}$ and $\circled{C}$.
Increasing temperature, we turn on e-e scattering, which helps to redistribute the momentum as long as the boundary scattering dominates, increasing thus the resistivity.
However, above the temperature $\circled{2}$, when the e-e scattering length $l_\mathrm{ee}$ gets shorter than the width of the channel, electrons cannot reach the walls anymore and scatter on each other, so that only a small percentage of electrons situated in the vicinity of the boundary can reach it.
This is the conventional hydrodynamical regime $\circled{E}$: the increasing of the temperature shortens $l_\mathrm{ee}$ and thus decreases the viscosity.
This behavior obviously lowers the net resistivity. 
The decrease of the resistivity lasts until the dissipative term originating from the impurity scattering starts being relevant (line $\circled{5}$).
After this point the behavior of the curve retraces the one in Fig.~\ref{fig:cross}a):
We enter the diffusive regime $\circled{A}$ characterized by a constant resistivity, which depends neither on temperature nor on chemical potential.
Afterwards, crossing $\circled{4}$, we are in the pseudo\-diffusive regime $\circled{G}$ with increasing resistivity with temperature.
The crossover to the e-h scattering regime is not shown in the plot~\ref{fig:cross}b), since a very low value of $\vartheta$ pushes the line $\circled{7}$ to high temperatures.

The remaining plot~\ref{fig:cross}c) demonstrates at low temperatures the same features as plot~\ref{fig:cross}b).
The difference lies in the fact that the lines $\circled{2}$ and $\circled{4}$ are swapped.
This peculiarity opens a parameter window where the hydrodynamic approximation is still valid, but the Dirac system is close to charge neutrality.
This condition results in heat flow hydrodynamics, which has already been studied in graphene~\cite{Morpurgo2017}.
In this regime $\circled{F}$, the drop in the viscosity is compensated by the decrease of the ratio of the numbers of e-e and e-h scattering events (see discussion in Section~\ref{sec:regimes}). 
Crossing the line $\circled{4}$, the system enters the pseudo\-diffusive regime $\circled{G}$ straight away, developing similarly to the plot~\ref{fig:cross}b).
Comparing the main plots in Fig.~\ref{fig:cross}b) and~\ref{fig:cross}c), one may notice that despite the wider region of the hydrodynamical behavior in plot~\ref{fig:cross}c) it shows a less pronounced Gurzhi effect than plot~\ref{fig:cross}b), since the drop of the resistance happens only in the conventional Poiseuille regime $\circled{E}$, which is narrower in the plot~\ref{fig:cross}c).

\begin{figure}
\centering
\includegraphics[width=.78\columnwidth]{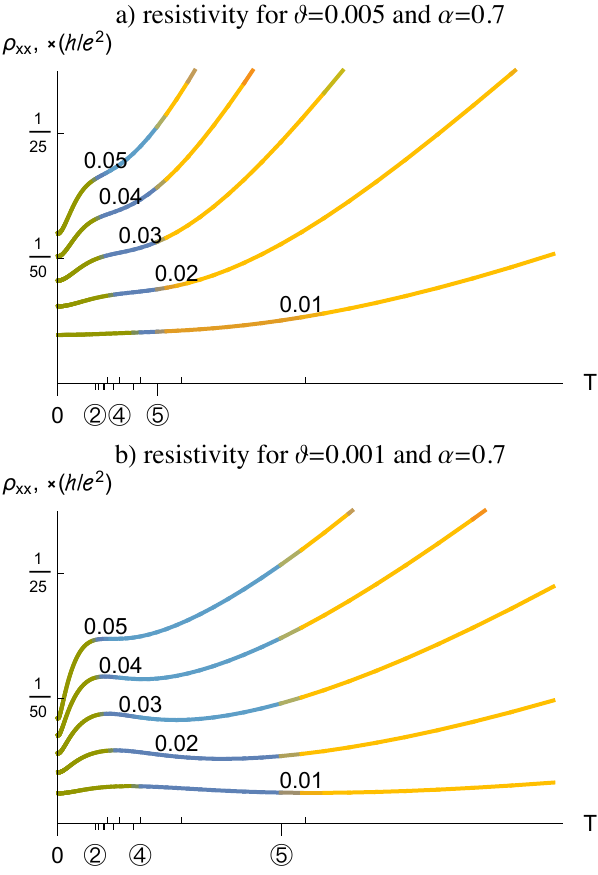}
\caption{Demonstration of the importance of the sample cleanliness for the Gurzhi effect.
The plots are drawn for different chemical potentials, under the choice $\alpha=0.7$, $\vartheta=10^{-3}$ [plot b)] and $\vartheta=5\times10^{-3}$ [plot a)].
The chemical potential is characterized by the parameter $\lambda_\mathrm{F}/W$, the value of which (running from $0.01$ to $0.05$) is written over each curve in the plots.
Short ticks on the $x$ axis pointing upwards correspond to the position of the line $\circled{4}$ for each curve, and short ticks pointing downwards---to the line $\circled{2}$.
The tick order repeats the reverse order of the values of the parameter $\lambda_\mathrm{F}/W$, i.e.\ the most left corresponds to $0.05$, next to $0.04$, etc.
The line $\circled{5}$ has the same position for the curves and is marked by a longer tick pointing downwards.
}
\label{fig:biases}
\end{figure}

As we mentioned before, we use a different value of $\vartheta$ in order to resolve the Poiseuille regions $\circled{E}$ and $\circled{F}$ demonstrating the enlargement of the hydrodynamic region at decreasing $\vartheta$.
It is important to stress that this is not the only effect of $\vartheta$: its low value is critical for the existence of the relativistic Gurzhi effect.
This statement is best illustrated by Fig.~\ref{fig:biases}.
The plots~\ref{fig:biases}a) and~\ref{fig:biases}b) are drawn for the same chemical potentials (characterized by the value $\lambda_\mathrm{F}/W=\hbar v/W\mu$ given for the each curve) and the same $\alpha=0.7$, but the cleanliness of the systems is different: $\vartheta=5.0\times10^{-3}$ and $\vartheta=1.0\times10^{-3}$, correspondingly.
Note that we are far away from the case illustrated by Fig.~\ref{fig:roadmap}c), since for both plots in Fig.~\ref{fig:biases}, $\alpha^{2}/\vartheta > 100$.
As one can see, even at very small $\vartheta=0.005$, despite the existing hydrodynamical regime [see Fig.~\ref{fig:biases}a)], the growth of the resistance is monotonic.
With the increase of the chemical potential (decrease of $\lambda_\mathrm{F}/W$) the density of states get increased and the decoupling of the flow from the electric field (which manifest itself in the resistance growth in the pseudo\-diffusive regime) affects the hydrodynamics less.
Evidently, with the increase of $\mu$, the hydrodynamic region shrinks and finally disappears before the resistivity curve can demonstrate a decline.
At $\vartheta=0.001$ the line $\circled{5}$, which denotes the end of the Poiseuille regime due to the dumping by the impurity scattering, is shifted to the right in plot~\ref{fig:biases}b) by a factor $5$ in comparison to the plot~\ref{fig:biases}a).
Weakening thus the influence of the pseudo\-diffusive regime, this shift allows the change of the $d^{2}V/dIdT$ sign before the Poiseuille regime interval collapses.

\section{Summary}
\label{sec:summary}

We have studied different mechanisms of electron transport in 2D Dirac materials within Boltzmann theory.
The interplay of electron-electron, impurity, and boundary scattering results in rich transport physics.
We have classified possible regimes and described the temperature and chemical potential dependencies of the resistivity for each of those.
We have put the focus on the hydrodynamic behavior in the channel configuration of the system. 
Two different hydrodynamic (Poisseuille) regimes are found in a narrow range of temperature and chemical potential.
These are manifestations of the relativistic Gurzhi effect.
The size of the hydrodynamic range depends on the cleanliness of the sample.
We have pointed out an existence of an additional pseudodiffusive regime, which can be realized for $\mu\ll T$. 
A quantitative numerical solution of the Boltzmann equation is provided, confirming the qualitative estimations of the resistivity.
Numerical computation also demonstrates that the signature of the Poisseuille regime, in the presence of weak disorder $\vartheta \ll \alpha^{2}$, does not always guarantee the phenomenology of the relativistic Gurzhi effect.  
We have shown that the existence of the nonmotonicity of the resistivity as a function of temperature is very sensitive to disorder and reveals itself only below some critical value of the disorder strength.

\begin{acknowledgments}
Financial support by the DFG (SPP1666 and SFB1170 "ToCoTronics") and the Helmholtz Foundation (VITI) is gratefully acknowledged.
We thank  I.~Gornyi, M.~Titov, B.~Narozhny, and M.~Kharitonov for discussions.
\end{acknowledgments}

\appendix

\section{Confinement and boundary conditions for 2D Dirac electrons}
\label{apx:dconf}

The confinement of excitations described by the 2D Dirac equation can be realized due to the unused third Pauli matrix, which allows us to open a gap outside the confinement area.
For example, all particles with an energy $\epsilon<M$ cannot propagate at $y>0$ if we construct the Hamiltonian~\cite{Berry1987}
\begin{equation*}
H = \sigma^{x}p_{x} + \sigma^{y}p_{y} + \sigma^{z} M \theta(y),
\end{equation*}
where we put $v=1$ for simplicity.
Taking the limit of $M\to\infty$ we get an effective boundary condition which preserves the chirality~\cite{Kharitonov2017}.
A more general boundary condition can, in principle, be chiral asymmetric and still satisfy the requirement of zero current perpendicular to the boundary.
The general boundary condition for the spinor $(\psi_{\uparrow},\psi_{\downarrow})$ on the boundary looks like
\begin{equation*}
c_{\uparrow}\psi_{\uparrow} + c_{\downarrow}\psi_{\downarrow} = 0,
\quad
c_{\alpha}=
\begin{cases}
e^{i\varphi/2}\cos\theta_{b} & \alpha=\uparrow, \\
e^{-i\varphi/2}\sin\theta_{b} & \alpha=\downarrow,
\end{cases}
\end{equation*}
where $\varphi$ is the orientation of the boundary in the $x$--$y$ coordinate system ($\varphi=0$ if the boundary is parallel to the $x$ axis).
The parameter $\theta_{b}$ encodes the microscopic physics of the boundary.
At some values of $\theta_{b}$ an additional 1D edge state can develop~\cite{Ostaay2011,Kharitonov2017}.
The net solution at energy $\epsilon=\sqrt{p_{x}^{2}+p_{y}^{2}}$ can be written as
\begin{equation*}
\psi= A e^{ip_{x}x + ip_{y}y} + B e^{ip_{x}x - ip_{y}y} + C e^{iqx-\sqrt{q^{2}-\epsilon^{2}}y},
\end{equation*}
where $q>\epsilon>p_{x}$, $|A|=|B|$ in order to cancel out the current perpendicular to the boundary.
The amplitude $C$ denotes the contribution of the edge mode.

\section{Electron-electron scattering thermalization}
\label{apx:eethem}

The collision integral for e-e scattering has a complicated non\-linear functional dependence on the distribution function $f$:
\begin{multline*}
\mathcal{I}_\mathrm{ee}[f_{0}]= \sum_{123} w_{0123}
[f_{0}(1-f_{1})f_{2}(1-f_{3}) -\\- (1-f_{0})f_{1}(1-f_{2})f_{3}] 
\prod_{i}\delta(q_{0}^{(i)}-q_{1}^{(i)}+q_{2}^{(i)}-q_{3}^{(i)}).
\end{multline*}
In order to achieve a detailed balance and set the collision integral to zero, we need that distribution function obeys
\begin{equation*}
f_{0}(1-f_{1})f_{2}(1-f_{3}) = (1-f_{0})f_{1}(1-f_{2})f_{3}
\end{equation*}
for every set of the states $\breve{\mathbf{p}}_{0/1/2/3}$.
These states are not arbitrary, but are restricted by the conservation laws [the laws are enumerated by an index $(i)$], i.e.,
\begin{equation*}
q_{0}^{(i)}+q_{2}^{(i)} = q_{1}^{(i)}+q_{3}^{(i)}.
\end{equation*}
Note that the detailed balance condition for the distribution function can be rewritten as
\begin{equation*}
h_{0}+h_{2}=h_{1}+h_{3},
\quad\text{where}\quad
h_{n}=\log \frac{f_{n}}{1-f_{n}}.
\end{equation*}
Thus, we can satisfy the detailed balance condition using the ansatz $h_{n}=\sum_{i} \beta^{(i)} q_{n}^{(i)}$, where $\beta^{(i)}$ are constants independent on the states $\breve{\mathbf{p}}_{n}$.
Inverting the definition of the value $h$, i.e., $f=(e^{h}+1)^{-1}$, we derive the general distribution function that sets the e-e collision integral to zero
\begin{equation*}
f_{n}= \left[\mathrm{exp}\left( \sum_{i} \beta^{(i)} q_{n}^{(i)} \right)+1 \right]^{-1}.
\end{equation*}
In the case when the conserving parameters $q^{(i)}$ are energy $\epsilon$, components of the momentum $p_{x}$ and $p_{y}$, and the number of particles (then $q^{(n.p.)}=1$), we get
\begin{equation*}
f= \frac{1}{e^{\beta \epsilon + \beta^{(m.x)} p_{x} + \beta^{(m.y)}p_{y} + \beta^{(m.p.)} }+1},
\end{equation*}
where $\beta=1/T$, $\beta^{(m.x/y)}=-V_{x/y}/T$, and $\beta^{(m.p.)}=-\mu/T$.
This expression is equivalent to the function $f_\mathrm{V}$ in Eq.~\eqref{eq:driftv}.

\section{Minimal impurity scattering}

\label{apx:limax}
In realistic experimental setups, however, the limit $l_\mathrm{i}\to \infty$ cannot be achieved by decreasing temperature $T$ and chemical potential $\mu$.
Due to various reasons, such as inelastic scatterers, localized impurity states, electron and hole puddles, etc., the impurity scattering length has typically an upper limit $l_{i0}$.
The impurity scattering length can thus be approximated as $l_\mathrm{i}^{-1}=l_{i0}^{-1}+\vartheta |\epsilon|/\hbar v$.
Despite that the formalism used in this work allows us to implement this effect, in order to simplify the physical picture, we assume that this maximal length (that corresponds to the minimal scattering) is much larger than all other scattering lengths like channel width $W$ and electron-electron scattering length $l_\mathrm{ee}$.
Thus, we neglect it in the discussion of different transport regimes.

\section{Collinear scattering}
\label{apx:collin}

In general, the prefactor in the expansion of the collision integral over $\delta f=f-f_\text{V}$ is momentum dependent, i.e. $\mathcal{I}[f]=\sum_{\mathbf{p}'}R_{\mathbf{p},\mathbf{p}'}(\delta f_{\mathbf{p}'}-\delta f_{\mathbf{p}})$.
In case of a smooth dependence of $R_{\mathbf{p},\mathbf{p}'}$ on the momentum $\mathbf{p}'$, the first term can be neglected and the collision integral will take the form of Eq.~\eqref{eq:Iee}, where $\tau_\text{ee}^{-1}=\sum_{\mathbf{p}'}R_{\mathbf{p},\mathbf{p}'}$.
The collinear scattering in Dirac systems is enhanced due to the linearity of the spectrum and manifests itself in a logarithmically large value of $R_{\mathbf{p},\mathbf{p}}$.
This enhancement comes from the $1/|pp'-\mathbf{p}\cdot\mathbf{p}'|$ divergence of the scattering amplitude $R_{\mathbf{p},\mathbf{p}'}$, which reveals itself if the group velocities of initial and finite states are collinear~\cite{Fritz2008}.
The corresponding rate $\tau_\mathrm{cee}^{-1}\propto R_{\mathbf{p},\mathbf{p}}$ can be estimated as $\tau_\mathrm{cee}^{-1}\sim \tau_\mathrm{ee}^{-1}\log(\alpha^{-2})$. 
Note that for $\alpha\sim1$ the additional collinear scattering integral is not just added to $\mathcal{I}_\text{ee}$, but simply drops out, since $R$ acquires back its smooth dependence on the momenta. 

In the collinear limit, the energy and momentum conservation laws almost coincide, since the momenta are $\mathbf{p}_{j}\approx v^{-1}\epsilon_{j}\mathbf{n}$.
The collinear scattering collision integral should also preserve the momentum and the particle number.
This allows us to use the Callaway ansatz idea for the collision integral with the modification that we let the parameters of $f_\mathrm{V}$ [including $\mu$, playing the role of the chemical potential, see Eq.~\eqref{eq:driftv}] vary with the velocity direction $\mathbf{n}=\lambda\mathbf{p}/p$.
The corresponding collision integral can be phenomenologically written as
\begin{equation*}
\mathcal{I}_\mathrm{cee}[f] = \frac{f_{\mathrm{V},\mathbf{n}} - f_{\epsilon,\mathbf{n}}}{\tau_{cee}},
\end{equation*}
where
\begin{equation*}
f_{\mathrm{V},\mathbf{n}}=f_\mathrm{F}\bigl(\epsilon-(\epsilon-\mu)v^{-1}V_{\mathbf{n}} -\delta\mu_{\mathbf{n}}\bigr).
\end{equation*}
Performing the expansion of the distribution function $f$ in $\chi$, we can write the collision integral with the explicit expressions for the $\mathbf{V}_{\mathbf{n}}$ and $\delta\mu_{\mathbf{n}}$ as
\begin{equation}
\mathcal{I}_\mathrm{cee}[f] \approx -f_\mathrm{F}'\times \frac{1}{\tau_\mathrm{cee}}\bigl((\epsilon-\mu)v^{-1}V_{\mathbf{n}} +\delta\mu_{\mathbf{n}} - \chi_{\epsilon,\mathbf{n}} \bigr).
\label{eq:Icee}
\end{equation}
This equation is similar to Eq.~\eqref{eq:kinchi}, where $V_{\mathbf{n}}$ and $\delta\mu_{\mathbf{n}}$ are obtained from the particle and momentum conservation laws for each given direction, namely
\begin{equation*}
\int [f_{\mathrm{V},\mathbf{n}} - f_{\epsilon,\mathbf{n}}] |\epsilon|d\epsilon = 0, 
\qquad
\int [f_{\mathrm{V},\mathbf{n}} - f_{\epsilon,\mathbf{n}}] \epsilon|\epsilon|d\epsilon = 0.
\end{equation*}
The general expression $(\epsilon-\mu)v^{-1}V_{\mathbf{n}} +\delta\mu_{\mathbf{n}}$ in Eq.~\eqref{eq:Icee} is equivalent to the $\chi$ that we use in Appendix~\ref{apx:expansion}, where we justify the expansion of $f$ in $\chi$.
Higher orders in the expansion in $\epsilon-\mu$ do not play an important role here.
The symmetry of the function $\chi$ with respect to the e-h hole inversion, is encoded in the dependence on $\mathbf{n}$.
This symmetry is important in case of $\mu\ll T$ and can be extracted by the integrals $\overline{\vphantom{X}\ldots}$ and $\underline{\vphantom{X}\ldots}$ defined in Eqs.~\eqref{eq:defjol} and~\eqref{eq:defVul}, correspondingly.

In other words, when moving from Eq.~\eqref{eq:kineqfull} to Eq.~\eqref{eq:kinchi}, we imply that (a) the distribution $f$ is close to the hydrodynamic distribution $f_\mathrm{V}$, (b) it is almost equilibrated in each chosen direction with respect to the different energies scales, see Appendix~\ref{apx:expansion}.
The collinear scattering implies a different (intermediate) saturation value $f_{V,\mathbf{n}}$ of the distribution function (forced by $\mathcal{I}_\mathrm{cee}[f]$).
However, eventually, the distribution function $f_\mathrm{V}$ (forced by $\mathcal{I}_\mathrm{ee}[f]$) is reached at long time scales.

\section{Dirac fluid hydrodynamics}
\label{apx:didro}

The hydrodynamic equations for a Dirac system can be derived from the Boltzmann equation~\cite{Muller2009a,Briskot2015} 
\begin{equation*}
\dot{f}_{\lambda,\mathbf{p}} + v \mathbf{n}\cdot\nabla f_{\lambda,\mathbf{p}} + e\mathbf{E}\cdot\partial_{\mathbf{p}} f_{\lambda,\mathbf{p}}= I_\text{ee}[f]
\end{equation*}
multiplying it by the conserved variables (for the momentum this means that $\sum_{\pm,\mathbf{p}} \mathbf{p}I_\text{ee}[f] =0$ for any $f_{\lambda,\mathbf{p}}$) and integrating over all states.
The conserved quantity that results in the analog of the Navier-Stokes equation is the excitation momentum, and the equation takes form
\begin{equation}
\dot{\mathbf{P}} + \nabla\hat{\Pi} - e\mathbf{E} N = 0.
\label{eq:continuity}
\end{equation}
Net momentum density $\mathbf{P}$ (or energy current according to Ref.~\onlinecite{Narozhny2015}), 
charge (electrical) current $\mathbf{j}$, charge density $N$, and flow density tensor $\Pi$ are defined as
\begin{align*}
\mathbf{P} &=\sum_{\pm,\mathbf{p}} \mathbf{p} f_{\pm,\mathbf{p}},
&
\mathbf{j} &= ev\sum_{\pm,\mathbf{p}} \mathbf{n} f_{\pm,\mathbf{p}},
\\
N&=\sum_{\pm,\mathbf{p}} f_{\pm,\mathbf{p}},
&
\Pi_{ij}&=v\sum_{\pm,\mathbf{p}} n_{i}p_{j} f_{\pm,\mathbf{p}}.
\end{align*}
Here, we use a substitution $f_{-,\mathbf{p}}\to f_{-,\mathbf{p}}-1$ in order to avoid the divergency in summation over the valence band.
In the hydrodynamic limit, the collision integral is the dominating term in the kinetic equation, so the distribution function can be expanded around the shifted distribution function $f_\text{V}$ introduced in Eq.~\eqref{eq:driftv}, $f=f_\text{V}+\delta f$, since $I_\text{ee}[f_{V}]=0$.
Expanding the collision integral in $\delta f$, we get the Callaway ansatz [see Eq.~\eqref{eq:Iee}], and from the kinetic equation we obtain the non\-equilibrium correction to the distribution function $\delta f \approx  - \tau_\text{ee} v \mathbf{n}\cdot\nabla f_{V}$.
Calculating the macroscopic values up to the linear term in drift velocity $\mathbf{V}$ (we assume $V\ll v$) we get 
\begin{gather*}
\mathbf{P} =  M\mathbf{V},
\qquad
\mathbf{j} = eN\mathbf{V},
\\
\Pi_{ij}= \frac{M v^{2}}{3}  \delta_{ij} -  \zeta \bigl(\partial_{l}P_{l}\, \delta_{ij}+ \partial_{i}P_{j}+\partial_{j}P_{i}\bigr),
\end{gather*}
%
where $\zeta=v^{2} \tau_\mathrm{ee}/4$ is the kinematic viscosity, $N$ and $M$ are charge and ``mass'' densities, correspondingly:
\begin{equation*}
M = \frac{3}{2\pi\hbar^{2}v^{4}}F_{3}^{+},
\qquad
N = \frac{1}{2\pi\hbar^{2}v^{2}} F_{2}^{-}.
\end{equation*}
Here $F_{n}^{\pm}=-T^{n}[\mathrm{Li}_{n}(-e^{-\mu/T})\pm \mathrm{Li}_{n}(-e^{\mu/T})]$, $T$ is the temperature, $\mu$ the chemical potential, and $\mathrm{Li}_n(z)=\sum _{k=1}^{\infty } \frac{z^k}{k^n}$ the polylogarithmic function.
The charge density is connected with the number of electrons and holes as $N\propto |n_{e}-n_{h}|$, where
\begin{equation*}
n_{e} = \int f_{+,\mathbf{p}} \frac{d^{2}\mathbf{p}}{(2\pi\hbar)^{2}},
\quad\text{and}\quad
n_{h} = \int (1-f_{-,\mathbf{p}}) \frac{d^{2}\mathbf{p}}{(2\pi\hbar)^{2}}.
\end{equation*}
Substituting these expressions into the Eq.~\eqref{eq:continuity} we get the hydrodynamic equation (without the convection term, since it is quadratic in $\mathbf{V}$):
\begin{equation*}
\dot{\mathbf{P}} + \frac{v^{2}}{3} \mathrm{grad}\,M- \zeta
\left[
2\,\mathrm{grad}\,\mathrm{div}\,\mathbf{P} +\Delta\mathbf{P}
\right] + e\mathbf{E} N = 0.
\end{equation*}
For constant temperature and chemical potential, in the case of the stationary laminar flow, the equation gets simplified, yielding a well-known parabolic velocity profile:
\begin{equation*}
\partial^{2}_{y}V_{x} = eE N/\zeta M,
\qquad
V_{x} = (eE N/2\zeta M) (W-y)y.
\end{equation*}
Here, the boundary conditions which correspond to the zero boundary slip length~\cite{Torre2015} $V_{x}(0)=V_{x}(W)=0$ are applied.
Averaging the current density over the coordinate $y$ we obtain the conductivity
\begin{align*}
\frac{1}{\rho_{xx}} &= e^{2} N \frac{1}{W}\int V_{x}(y)dy \sim e^{2}\frac{W^{2}}{\zeta} \frac{N^{2}}{M} 
=\\&=
\frac{e^{2}}{\hbar^{2}v^{2}}\frac{W^{2}}{\tau_\mathrm{ee}} \times
\begin{cases}
1/\mu & \text{for $\mu\gg T$,}\\
\mu^{2}/T & \text{for $\mu\ll T$,}
\end{cases}
=\\&=
e^{2} \nu_\mathrm{eff} \,v\, l_\mathrm{eff} \times  
\begin{cases}
1& \text{for $\mu\gg T$,}\\
\mu^{2}/T^{2} & \text{for $\mu\ll T$,}
\end{cases}
\end{align*}
where $l_\mathrm{eff}=W^{2}/l_\mathrm{ee}$ and the density of states $\nu_\mathrm{eff}$ for Dirac systems is defined in Eq.~\eqref{eq:nueff}.
The obtained formula complies to the definition of the effective mean free path in Eq.~\eqref{eq:resistance}, and derives the factor $\mu^{2}/T^{2}$ for the Dirac system at $\mu\ll T$ given in Eq.~\eqref{eq:leffpP} explicitly.

%

\section{Expansion around Fermi surface}
\label{apx:expansion}

The general distribution function can be parametrized using the coordinate dependent chemical potential $\mu\equiv\mu(\mathbf{r})$ as
\begin{equation*}
f = f_\mathrm{F}(\epsilon+\chi) = \frac{1}{e^{[\epsilon_{\mathbf{p}}+\chi_{\mathbf{p}}(\mathbf{r})-\mu(\mathbf{r})]/T}+1},
\end{equation*}
where the function $\chi$ depends on momentum and coordinates.
We can always require that
\begin{equation*}
\left<\chi_{\mathbf{p}}(\mathbf{r}) \delta\bigl(\epsilon_{\mathbf{p}}-\mu(\mathbf{r})\bigr)\right>=0,
\end{equation*}
attributing a non\-zero average of $\chi$ to the chemical potential.
Let us check whether the expansion over $\chi$ is valid under the condition $|\partial_{\mathbf{p}}\chi|\ll v$, i.e.
\begin{equation*}
f_\mathrm{F}(\epsilon+\chi) \approx f_\mathrm{F}(\epsilon) + f_\mathrm{F}'(\epsilon)\,\chi,
\end{equation*}
where $f_\mathrm{F}'=\partial_{\epsilon}f_\mathrm{F}(\epsilon)$.
If $|\partial_{\mathbf{p}}\chi|\ll v$ then we can choose some arbitrary energy scale $\Delta$ that satisfies the condition
\begin{equation*}
|\chi|< \frac{|\partial_{\mathbf{p}}\chi|}{v} \Delta \ll T \ll \Delta.
\end{equation*}
In the interval $|\epsilon-\mu|<\Delta$, the function $\chi$ is limited by $|\chi|< |\partial_{\mathbf{p}}\chi|\,\Delta/v$.
Therefore, in this interval $\chi \ll T$ and the expansion is possible.
Outside this interval the expansion is still valid since
\begin{align*}
f_\mathrm{F}(\epsilon+\chi) \approx f_\mathrm{F}(\epsilon)&\approx \theta(\mu-\epsilon) + O[e^{-\Delta/T}],
\\
f_\mathrm{F}'(\epsilon)&\approx T^{-1}O[e^{-\Delta/T}],
\end{align*}
so that the value of the function $\chi$ is irrelevant due to the suppression by the small exponent.

\section{Symmetry of the distribution function}
\label{apx:symmetry}

A set of statements about the angular dependence of the function $\chi$ can be made by simple physical arguments and elementary analysis of Eq.~\eqref{eq:kinchi}.
The kinetic equation~\eqref{eq:kinchi} is linear in both $\chi$ and electric field term $\mathbf{E}\cdot\mathbf{n}$.
This automatically means that the solution has to be proportional to this term.
Introducing the angle $\varphi$ between the vector $\mathbf{n}$ and the electric field $\mathbf{E}$, we can use the parametrization $\chi = eE\cos\varphi\,l(\varphi)$, which we also introduce in the text in Eq.~\eqref{eq:condtl}.
If the electric field was directed in the opposite way, the distribution function would be reflected with respect to the line perpendicular to the electric field.
Mathematically this means
\begin{equation*}
\chi(\varphi) = - \chi(\pi-\varphi), \qquad\text{or}\qquad l(\varphi) = l(\pi-\varphi).
\end{equation*}
This relation automatically means that the zero harmonic in $\chi$ is absent $\langle\chi\rangle=0$, or using the parametrization in terms of the mean free path, $\langle\cos\varphi\, l(y,\epsilon,\varphi)\rangle=0$.

\section{Derivation of the equation for the effective mean free path}
\label{apx:leqn}

Using the parametrization $\chi = e E \cos\varphi\, l(y,\epsilon,\varphi)$ in Eq.~\eqref{eq:kinchi}, where the function $l\equiv l(y,\epsilon,\varphi)$ depends on the coordinate across the channel $y$, energy $\epsilon$, and angle $\varphi$, together with the fact that $\langle\chi\rangle=0$, which is derived in Appendix~\ref{apx:symmetry}, we obtain the equation 
\begin{equation}
\sin\varphi \, \partial_{y}l  - 1 = \frac{\vartheta}{2\hbar v} |\epsilon|\left( \tilde{l}- 2l \right) + \frac{1}{l_\mathrm{ee}}\left( \epsilon \,\underline{\tilde{l}}-l\right)
\label{eq:kinl}
\end{equation}
where $\tilde{l}\equiv\tilde{l}(y,\epsilon)$ depends on coordinate and energy, and $\underline{\tilde{l}}\equiv \underline{\tilde{l}}(y)$ on coordinate only:
\begin{equation*}
\tilde{l}=2 \langle l\,\cos^{2}\varphi \rangle, \qquad
\underline{\tilde{l}}\equiv \underline{\tilde{l}(y,\epsilon)}.
\end{equation*}
The underline notation is defined in Eq.~\eqref{eq:defVul}.
The boundary condition from Eq.~\eqref{eq:boundary} for fixed energy $\epsilon$ (since the boundary scattering is elastic) takes the form
\begin{subequations}
\begin{align}
l(0,\epsilon,\varphi)\ &= \frac{1}{\pi\cos\varphi} \int_{\pi}^{2\pi} l(0,\epsilon,\varphi')\cos\varphi' d\varphi'
\nonumber\\&\hspace{3cm}
\text{for $0<\varphi<\pi$,} \\
l(W,\epsilon,\varphi) &= \frac{1}{\pi\cos\varphi}\int_{0}^{\pi} l(W,\epsilon,\varphi')\cos\varphi' d\varphi' 
\nonumber\\&\hspace{3cm}
\text{for $\pi<\varphi<2\pi$.}  
\end{align}
\label{eq:bcl}
\end{subequations}
The conductivity in these terms can be written as
\begin{equation}
\sigma = \frac{e^{2}}{2\pi \hbar} \frac{\mu}{\hbar v W}\int_{0}^{W}  \overline{\,\tilde{l}\,\,}dy,
\label{eq:condl}
\end{equation}
where overbar is defined in Eq.~\eqref{eq:defjol}.
The solution of the kinetic equation~\eqref{eq:kinl} with boundary conditions in Eqs.~\eqref{eq:bcl} is
\begin{subequations}
\begin{align}
l(y,\epsilon,\varphi) &= 
\int_{0}^{y} \Biggl(1+\frac{\vartheta}{2\hbar v} |\epsilon| \tilde{l}(y',\epsilon) + \frac{\epsilon}{l_\mathrm{ee}}\underline{\tilde{l}}(y')\Biggr)
\nonumber\\ &\hspace{1cm}\times
\frac{e^{-\frac{y-y'}{l_{tot}\sin\varphi}}}{\sin\varphi} dy',
\text{ for }\varphi\in[0,\pi],
\\
l(y,\epsilon,\varphi) &= 
\int_{y}^{W} \Biggl(1+\frac{\vartheta}{2\hbar v} |\epsilon| \tilde{l}(y',\epsilon) + \frac{\epsilon}{l_\mathrm{ee}}\underline{\tilde{l}}(y')\Biggr)
\nonumber\\ &\hspace{1cm}\times
\frac{e^{-\frac{y'-y}{l_{tot}|\sin\varphi|}}}{|\sin\varphi|} dy',
\text{ for }\varphi\in[\pi,2\pi],
\end{align}
\label{eq:soll}
\end{subequations}
where
\begin{equation*}
\frac{1}{l_{tot}} = \frac{\vartheta}{\hbar v} |\epsilon| + \frac{1}{l_\mathrm{ee}}.
\end{equation*}
The full angle-resolved information about the distribution is not needed for the calculation of the conductivity in Eq.~\eqref{eq:condl}.
Therefore, we derive the equation for $\tilde{l}$ by multiplying Eqs.~\eqref{eq:soll} by $\cos^{2}\varphi$ and integrating them over all angles:
\begin{multline}
\tilde{l}(y,\epsilon) = 
\int_{0}^{W}\left(1+\frac{\vartheta}{2\hbar v} |\epsilon| \tilde{l}(y',\epsilon) + \frac{\epsilon}{l_\mathrm{ee}}\underline{\tilde{l}}(y')\right)\\\times
K(|y-y'|/l_{tot}) dy',
\end{multline}
where
\begin{equation*}
K(z) = \frac{2}{\pi} \int_{0}^{\pi/2} \frac{\cos^{2}\varphi}{\sin\varphi} e^{-z/\sin\varphi}d\varphi.
\end{equation*}
The function $K$ can be alternatively defined through the integral $K(z) = \frac{2}{\pi} \int_{1}^{\infty} t^{-2}\sqrt{t^{2}-1} e^{-zt}dt$, relating it to the modified Bessel function of the second kind $K_{n}(z)$ as $\pi z\,\partial^{2}_{z}K(z) = 2 K_{1}(z)$.

\section{Numerical approach}
\label{apx:numerical}

The function $\tilde{l}$ as a function of the coordinate $y$ exists only in the interval $[0,W]$, so that the Fourier transform gives us a discrete but infinite set of the coefficients.
Since the distribution function is mirror symmetric with respect to the middle of the channel, $\tilde{l}$ is symmetric too: $\tilde{l}(W-y)=\tilde{l}(y)$, and all terms containing a $\sin$ in the Fourier expansion drop out:
\begin{equation}
\begin{split}
&\tilde{l}(y)=\sum_{n=0}^{\infty} \tilde{l}_{n} \cos\frac{2\pi n y}{W},
\qquad
\tilde{l}_{0}=\frac{1}{W}\int_{0}^{W} \tilde{l}(y)dy,
\\
&\tilde{l}_{n}=\frac{2}{W}\int_{0}^{W} \tilde{l}(y)\cos\frac{2\pi n y}{W}dy.
\end{split}
\label{eq:FourierExpansion}
\end{equation}
Using the notation $a=\frac{\vartheta}{2\hbar v} |\epsilon|$ and $b=\frac{\epsilon}{l_\mathrm{ee}}$ to make the expressions shorter, we can write the Fredholm equation in Eq.~\eqref{eq:tlK} as
\begin{equation*}
(1+\delta_{n0})\tilde{l}_{n} = K_{n0} + \sum_{m=0}^{\infty} \left(1+a \tilde{l}_{m} + b\underline{\tilde{l}_{m}}\right) K_{mn},
\end{equation*}
where
\begin{equation*}
K_{nm} = \frac{2}{W}\iint_{0}^{w} K\left(\frac{|y-y'|}{l_{tot}}\right)\cos\frac{2\pi n y}{W}\cos\frac{2\pi m y'}{W}dydy'.
\end{equation*}
The conductivity is proportional to the zero harmonic of the expansion~\eqref{eq:FourierExpansion}
\begin{equation}
1/\rho_{xx}= e^{2}\nu_{F}v_{F} \mu \overline{\,\tilde{l}_{0}\,}.
\label{eq:condtlmatr}
\end{equation}
Defining the infinite column $1_{0} = (1,0,0,\ldots)^{T}$ and diagonal matrix $U_{nm}=\delta_{m0}\delta_{n0}+\delta_{nm}$, we can write the equation in a matrix representation
\begin{equation}
U\tilde{l}=K\left(1_{0}+a\,\tilde{l}+b\,\underline{\tilde{l}}\right),
\label{eq:matrtl0}
\end{equation}
where $U$ and $K$ are square matrices of infinite dimensions.
Introducing the matrix $Q=(U-aK)^{-1}K$, Eq.~\eqref{eq:matrtl0} can be written as
\begin{equation*}
\tilde{l}=Q\left(1_{0}+b\,\underline{\tilde{l}}\right).
\end{equation*}
Integrating it over the energy, as defined in Eq.~\eqref{eq:defVul} for the underline symbol, we get the result
\begin{equation*}
\underline{\tilde{l}} = \left(1-\underline{bQ}\right)^{-1}\underline{Q}\,1_{0}.
\end{equation*}
Substituting it back into the equation for $\tilde{l}$, we obtain the final expression
\begin{equation*}
\tilde{l}=Q\left(1+b\left(1-\underline{bQ}\right)^{-1}
\underline{Q}\right)1_{0}.
\end{equation*}
Since we need only the first element of the column $\tilde{l}$ to calculate the conductivity [see Eq.~\eqref{eq:condtlmatr}], the formal result for the conductivity is
\begin{equation}
1/\rho_{xx}= e^{2}\nu_{F}v_{F} \mu \times
\overline{\,\left[Q\left(1+b\left(1-\underline{bQ}\right)^{-1}
\underline{Q}\right)\right]_{00}}.
\end{equation}
In order to calculate this formal expression it is sufficient to cut off the infinite matrices to a finite size.
Due to the rapid decay of the matrix $K_{nm}$ with large indices $n$ and $m$, as the numerical computation shows, the size $4\times4$ of the matrices is sufficient.

\section{Notations}
\label{apx:notes}

\begin{itemize}[leftmargin=35pt, parsep=-5pt, align=left, style=sameline]
\item[$I$] current along the sample
\item[$V$] voltage along the sample
\item[$\rho_{xx}$] resistivity
\item[$\mathbf{p}$] momentum
\item[$\breve{\mathbf{p}}$] full set of quantum numbers defining a quantum state (momentum and band number for Dirac spectra)
\item[$f,f_{\breve{\mathbf{p}}}$] distribution function
\item[$\epsilon_{\breve{\mathbf{p}}}$] dispersion relation
\item[$\mathbf{v}, v$] group velocity (vector, scalar)
\item[$\mathbf{n}$] electron propagation direction $\mathbf{v}/v$
\item[$\nu$] density of states
\item[$\mathcal{I}_{\ldots}{[f]}$] collision integral of the $\ldots$ scattering process
\item[$\tau^{-1}_{\ldots}(\breve{\mathbf{p}})$] the rate of the $\ldots$ scattering process for the electron in the state $\breve{\mathbf{p}}$
\item[$l_{\ldots}$] scattering length of the $\ldots$ scattering process
\item[$l_\mathrm{eff}$] effective mean free path 
\item[$L\times W$] length$\times$width of the channel
\item[$(x,y)$] coordinates along and across the channel
\item[$n_\mathrm{i}$] impurities density
\item[$U_{0}$] impurity strength
\item[$\hbar$] reduced Planck's constant
\item[$\mathbf{V}$] drift/flow velocity
\item[$\mu$] chemical potential
\item[$T$] electron temperature
\item[$\alpha$] e-e scattering strength (effective fine-structure constant)
\item[$\vartheta$] impurity scattering strength
\item[$\zeta$] kinematic viscosity
\item[$w$] e-e scattering probability
\end{itemize}

\bibliographystyle{apsrev4-1}
\bibliography{diracchannel}

\begin{thebibliography}{43}%
\makeatletter
\providecommand \@ifxundefined [1]{%
 \@ifx{#1\undefined}
}%
\providecommand \@ifnum [1]{%
 \ifnum #1\expandafter \@firstoftwo
 \else \expandafter \@secondoftwo
 \fi
}%
\providecommand \@ifx [1]{%
 \ifx #1\expandafter \@firstoftwo
 \else \expandafter \@secondoftwo
 \fi
}%
\providecommand \natexlab [1]{#1}%
\providecommand \enquote  [1]{``#1''}%
\providecommand \bibnamefont  [1]{#1}%
\providecommand \bibfnamefont [1]{#1}%
\providecommand \citenamefont [1]{#1}%
\providecommand \href@noop [0]{\@secondoftwo}%
\providecommand \href [0]{\begingroup \@sanitize@url \@href}%
\providecommand \@href[1]{\@@startlink{#1}\@@href}%
\providecommand \@@href[1]{\endgroup#1\@@endlink}%
\providecommand \@sanitize@url [0]{\catcode `\\12\catcode `\$12\catcode
  `\&12\catcode `\#12\catcode `\^12\catcode `\_12\catcode `\%12\relax}%
\providecommand \@@startlink[1]{}%
\providecommand \@@endlink[0]{}%
\providecommand \url  [0]{\begingroup\@sanitize@url \@url }%
\providecommand \@url [1]{\endgroup\@href {#1}{\urlprefix }}%
\providecommand \urlprefix  [0]{URL }%
\providecommand \Eprint [0]{\href }%
\providecommand \doibase [0]{http://dx.doi.org/}%
\providecommand \selectlanguage [0]{\@gobble}%
\providecommand \bibinfo  [0]{\@secondoftwo}%
\providecommand \bibfield  [0]{\@secondoftwo}%
\providecommand \translation [1]{[#1]}%
\providecommand \BibitemOpen [0]{}%
\providecommand \bibitemStop [0]{}%
\providecommand \bibitemNoStop [0]{.\EOS\space}%
\providecommand \EOS [0]{\spacefactor3000\relax}%
\providecommand \BibitemShut  [1]{\csname bibitem#1\endcsname}%
\let\auto@bib@innerbib\@empty
\bibitem [{\citenamefont {Molenkamp}\ and\ \citenamefont
  {de~Jong}(1994)}]{Molenkamp1994a}%
  \BibitemOpen
  \bibfield  {author} {\bibinfo {author} {\bibfnamefont {L.~W.}\ \bibnamefont
  {Molenkamp}}\ and\ \bibinfo {author} {\bibfnamefont {M.~J.~M.}\ \bibnamefont
  {de~Jong}},\ }\href {\doibase 10.1016/0038-1101(94)90244-5} {\bibfield
  {journal} {\bibinfo  {journal} {Solid-State Electronics}\ }\textbf {\bibinfo
  {volume} {37}},\ \bibinfo {pages} {551} (\bibinfo {year} {1994})}\BibitemShut
  {NoStop}%
\bibitem [{\citenamefont {de~Jong}\ and\ \citenamefont
  {Molenkamp}(1995)}]{Jong1995}%
  \BibitemOpen
  \bibfield  {author} {\bibinfo {author} {\bibfnamefont {M.~J.~M.}\
  \bibnamefont {de~Jong}}\ and\ \bibinfo {author} {\bibfnamefont {L.~W.}\
  \bibnamefont {Molenkamp}},\ }\href {\doibase 10.1103/PhysRevB.51.13389}
  {\bibfield  {journal} {\bibinfo  {journal} {Phys. Rev. B}\ }\textbf {\bibinfo
  {volume} {51}},\ \bibinfo {pages} {13389} (\bibinfo {year}
  {1995})}\BibitemShut {NoStop}%
\bibitem [{\citenamefont {Gurzhi}(1963)}]{Gurzhi1963}%
  \BibitemOpen
  \bibfield  {author} {\bibinfo {author} {\bibfnamefont {R.~N.}\ \bibnamefont
  {Gurzhi}},\ }\href {http://www.jetp.ac.ru/cgi-bin/e/index/e/17/2/p521?a=list}
  {\bibfield  {journal} {\bibinfo  {journal} {Sov. Phys. JETP}\ }\textbf
  {\bibinfo {volume} {17}},\ \bibinfo {pages} {521} (\bibinfo {year} {1963})},\
  \bibinfo {note} {[ZhETF, {\bf 44}, No. 2, 771, (1963)]}\BibitemShut {NoStop}%
\bibitem [{\citenamefont {Gurzhi}(1964)}]{Gurzhi1964}%
  \BibitemOpen
  \bibfield  {author} {\bibinfo {author} {\bibfnamefont {R.~N.}\ \bibnamefont
  {Gurzhi}},\ }\href {http://www.jetp.ac.ru/cgi-bin/e/index/e/19/2/p490?a=list}
  {\bibfield  {journal} {\bibinfo  {journal} {Sov. Phys. JETP}\ }\textbf
  {\bibinfo {volume} {19}},\ \bibinfo {pages} {490} (\bibinfo {year} {1964})},\
  \bibinfo {note} {[ZhETF, {\bf 46}, No. 2, 719, (1964)]}\BibitemShut {NoStop}%
\bibitem [{\citenamefont {Gurzhi}(1968)}]{Gurzhi1968}%
  \BibitemOpen
  \bibfield  {author} {\bibinfo {author} {\bibfnamefont {R.~N.}\ \bibnamefont
  {Gurzhi}},\ }\href
  {http://iopscience.iop.org/article/10.1070/PU1968v011n02ABEH003815/meta}
  {\bibfield  {journal} {\bibinfo  {journal} {Sov. Phys. Usp.}\ }\textbf
  {\bibinfo {volume} {11}},\ \bibinfo {pages} {255} (\bibinfo {year} {1968})},\
  \bibinfo {note} {[UPhN, {\bf 94}, 689, (1968)]}\BibitemShut {NoStop}%
\bibitem [{\citenamefont {Moll}\ \emph {et~al.}(2016)\citenamefont {Moll},
  \citenamefont {Kushwaha}, \citenamefont {Nandi}, \citenamefont {Schmidt},\
  and\ \citenamefont {Mackenzie}}]{Moll2016}%
  \BibitemOpen
  \bibfield  {author} {\bibinfo {author} {\bibfnamefont {P.~J.~W.}\
  \bibnamefont {Moll}}, \bibinfo {author} {\bibfnamefont {P.}~\bibnamefont
  {Kushwaha}}, \bibinfo {author} {\bibfnamefont {N.}~\bibnamefont {Nandi}},
  \bibinfo {author} {\bibfnamefont {B.}~\bibnamefont {Schmidt}}, \ and\
  \bibinfo {author} {\bibfnamefont {A.~P.}\ \bibnamefont {Mackenzie}},\ }\href
  {\doibase 10.1126/science.aac8385} {\bibfield  {journal} {\bibinfo  {journal}
  {Science}\ }\textbf {\bibinfo {volume} {351}},\ \bibinfo {pages} {1061}
  (\bibinfo {year} {2016})}\BibitemShut {NoStop}%
\bibitem [{\citenamefont {Bandurin}\ \emph {et~al.}(2016)\citenamefont
  {Bandurin}, \citenamefont {Torre}, \citenamefont {Kumar}, \citenamefont
  {Ben~Shalom}, \citenamefont {Tomadin}, \citenamefont {Principi},
  \citenamefont {Auton}, \citenamefont {Khestanova}, \citenamefont {Novoselov},
  \citenamefont {Grigorieva}, \citenamefont {Ponomarenko}, \citenamefont
  {Geim},\ and\ \citenamefont {Polini}}]{Bandurin2016}%
  \BibitemOpen
  \bibfield  {author} {\bibinfo {author} {\bibfnamefont {D.~A.}\ \bibnamefont
  {Bandurin}}, \bibinfo {author} {\bibfnamefont {I.}~\bibnamefont {Torre}},
  \bibinfo {author} {\bibfnamefont {R.~K.}\ \bibnamefont {Kumar}}, \bibinfo
  {author} {\bibfnamefont {M.}~\bibnamefont {Ben~Shalom}}, \bibinfo {author}
  {\bibfnamefont {A.}~\bibnamefont {Tomadin}}, \bibinfo {author} {\bibfnamefont
  {A.}~\bibnamefont {Principi}}, \bibinfo {author} {\bibfnamefont {G.~H.}\
  \bibnamefont {Auton}}, \bibinfo {author} {\bibfnamefont {E.}~\bibnamefont
  {Khestanova}}, \bibinfo {author} {\bibfnamefont {K.~S.}\ \bibnamefont
  {Novoselov}}, \bibinfo {author} {\bibfnamefont {I.~V.}\ \bibnamefont
  {Grigorieva}}, \bibinfo {author} {\bibfnamefont {L.~A.}\ \bibnamefont
  {Ponomarenko}}, \bibinfo {author} {\bibfnamefont {A.~K.}\ \bibnamefont
  {Geim}}, \ and\ \bibinfo {author} {\bibfnamefont {M.}~\bibnamefont
  {Polini}},\ }\href {\doibase 10.1126/science.aad0201} {\bibfield  {journal}
  {\bibinfo  {journal} {Science}\ }\textbf {\bibinfo {volume} {351}},\ \bibinfo
  {pages} {1055} (\bibinfo {year} {2016})}\BibitemShut {NoStop}%
\bibitem [{\citenamefont {Crossno}\ \emph {et~al.}(2016)\citenamefont
  {Crossno}, \citenamefont {Shi}, \citenamefont {Wang}, \citenamefont {Liu},
  \citenamefont {Harzheim}, \citenamefont {Lucas}, \citenamefont {Sachdev},
  \citenamefont {Kim}, \citenamefont {Taniguchi}, \citenamefont {Watanabe},
  \citenamefont {Ohki},\ and\ \citenamefont {Fong}}]{Crossno2016}%
  \BibitemOpen
  \bibfield  {author} {\bibinfo {author} {\bibfnamefont {J.}~\bibnamefont
  {Crossno}}, \bibinfo {author} {\bibfnamefont {J.~K.}\ \bibnamefont {Shi}},
  \bibinfo {author} {\bibfnamefont {K.}~\bibnamefont {Wang}}, \bibinfo {author}
  {\bibfnamefont {X.}~\bibnamefont {Liu}}, \bibinfo {author} {\bibfnamefont
  {A.}~\bibnamefont {Harzheim}}, \bibinfo {author} {\bibfnamefont
  {A.}~\bibnamefont {Lucas}}, \bibinfo {author} {\bibfnamefont
  {S.}~\bibnamefont {Sachdev}}, \bibinfo {author} {\bibfnamefont
  {P.}~\bibnamefont {Kim}}, \bibinfo {author} {\bibfnamefont {T.}~\bibnamefont
  {Taniguchi}}, \bibinfo {author} {\bibfnamefont {K.}~\bibnamefont {Watanabe}},
  \bibinfo {author} {\bibfnamefont {T.~A.}\ \bibnamefont {Ohki}}, \ and\
  \bibinfo {author} {\bibfnamefont {K.~C.}\ \bibnamefont {Fong}},\ }\href
  {\doibase 10.1126/science.aad0343} {\bibfield  {journal} {\bibinfo  {journal}
  {Science}\ }\textbf {\bibinfo {volume} {351}},\ \bibinfo {pages} {1058}
  (\bibinfo {year} {2016})}\BibitemShut {NoStop}%
\bibitem [{\citenamefont {Nam}\ \emph {et~al.}(2017)\citenamefont {Nam},
  \citenamefont {Ki}, \citenamefont {Soler-Delgado},\ and\ \citenamefont
  {Morpurgo}}]{Morpurgo2017}%
  \BibitemOpen
  \bibfield  {author} {\bibinfo {author} {\bibfnamefont {Y.}~\bibnamefont
  {Nam}}, \bibinfo {author} {\bibfnamefont {D.-K.}\ \bibnamefont {Ki}},
  \bibinfo {author} {\bibfnamefont {D.}~\bibnamefont {Soler-Delgado}}, \ and\
  \bibinfo {author} {\bibfnamefont {A.~F.}\ \bibnamefont {Morpurgo}},\ }\href
  {\doibase 10.1038/nphys4218} {\bibfield  {journal} {\bibinfo  {journal} {Nat
  Phys}\ }\textbf {\bibinfo {volume} {13}},\ \bibinfo {pages} {1207} (\bibinfo
  {year} {2017})}\BibitemShut {NoStop}%
\bibitem [{\citenamefont {Hancock}\ \emph {et~al.}(2011)\citenamefont
  {Hancock}, \citenamefont {van Mechelen}, \citenamefont {Kuzmenko},
  \citenamefont {van~der Marel}, \citenamefont {Br\"une}, \citenamefont
  {Novik}, \citenamefont {Astakhov}, \citenamefont {Buhmann},\ and\
  \citenamefont {Molenkamp}}]{Hancock2011}%
  \BibitemOpen
  \bibfield  {author} {\bibinfo {author} {\bibfnamefont {J.~N.}\ \bibnamefont
  {Hancock}}, \bibinfo {author} {\bibfnamefont {J.~L.~M.}\ \bibnamefont {van
  Mechelen}}, \bibinfo {author} {\bibfnamefont {A.~B.}\ \bibnamefont
  {Kuzmenko}}, \bibinfo {author} {\bibfnamefont {D.}~\bibnamefont {van~der
  Marel}}, \bibinfo {author} {\bibfnamefont {C.}~\bibnamefont {Br\"une}},
  \bibinfo {author} {\bibfnamefont {E.~G.}\ \bibnamefont {Novik}}, \bibinfo
  {author} {\bibfnamefont {G.~V.}\ \bibnamefont {Astakhov}}, \bibinfo {author}
  {\bibfnamefont {H.}~\bibnamefont {Buhmann}}, \ and\ \bibinfo {author}
  {\bibfnamefont {L.~W.}\ \bibnamefont {Molenkamp}},\ }\href {\doibase
  10.1103/PhysRevLett.107.136803} {\bibfield  {journal} {\bibinfo  {journal}
  {Phys. Rev. Lett.}\ }\textbf {\bibinfo {volume} {107}},\ \bibinfo {pages}
  {136803} (\bibinfo {year} {2011})}\BibitemShut {NoStop}%
\bibitem [{\citenamefont {B{\"u}ttner}\ \emph {et~al.}(2011)\citenamefont
  {B{\"u}ttner}, \citenamefont {Liu}, \citenamefont {Tkachov}, \citenamefont
  {Novik}, \citenamefont {Br{\"u}ne}, \citenamefont {Buhmann}, \citenamefont
  {Hankiewicz}, \citenamefont {Recher}, \citenamefont {Trauzettel},
  \citenamefont {Zhang},\ and\ \citenamefont {Molenkamp}}]{Buttner2011}%
  \BibitemOpen
  \bibfield  {author} {\bibinfo {author} {\bibfnamefont {B.}~\bibnamefont
  {B{\"u}ttner}}, \bibinfo {author} {\bibfnamefont {C.~X.}\ \bibnamefont
  {Liu}}, \bibinfo {author} {\bibfnamefont {G.}~\bibnamefont {Tkachov}},
  \bibinfo {author} {\bibfnamefont {E.~G.}\ \bibnamefont {Novik}}, \bibinfo
  {author} {\bibfnamefont {C.}~\bibnamefont {Br{\"u}ne}}, \bibinfo {author}
  {\bibfnamefont {H.}~\bibnamefont {Buhmann}}, \bibinfo {author} {\bibfnamefont
  {E.~M.}\ \bibnamefont {Hankiewicz}}, \bibinfo {author} {\bibfnamefont
  {P.}~\bibnamefont {Recher}}, \bibinfo {author} {\bibfnamefont
  {B.}~\bibnamefont {Trauzettel}}, \bibinfo {author} {\bibfnamefont {S.~C.}\
  \bibnamefont {Zhang}}, \ and\ \bibinfo {author} {\bibfnamefont {L.~W.}\
  \bibnamefont {Molenkamp}},\ }\href {\doibase 10.1038/nphys1914} {\bibfield
  {journal} {\bibinfo  {journal} {Nature Physics}\ }\textbf {\bibinfo {volume}
  {7}},\ \bibinfo {pages} {418} (\bibinfo {year} {2011})}\BibitemShut {NoStop}%
\bibitem [{\citenamefont {K{\"o}nig}\ \emph {et~al.}(2007)\citenamefont
  {K{\"o}nig}, \citenamefont {Wiedmann}, \citenamefont {Br{\"u}ne},
  \citenamefont {Roth}, \citenamefont {Buhmann}, \citenamefont {Molenkamp},
  \citenamefont {Qi},\ and\ \citenamefont {Zhang}}]{Konig2007}%
  \BibitemOpen
  \bibfield  {author} {\bibinfo {author} {\bibfnamefont {M.}~\bibnamefont
  {K{\"o}nig}}, \bibinfo {author} {\bibfnamefont {S.}~\bibnamefont {Wiedmann}},
  \bibinfo {author} {\bibfnamefont {C.}~\bibnamefont {Br{\"u}ne}}, \bibinfo
  {author} {\bibfnamefont {A.}~\bibnamefont {Roth}}, \bibinfo {author}
  {\bibfnamefont {H.}~\bibnamefont {Buhmann}}, \bibinfo {author} {\bibfnamefont
  {L.~W.}\ \bibnamefont {Molenkamp}}, \bibinfo {author} {\bibfnamefont {X.-L.}\
  \bibnamefont {Qi}}, \ and\ \bibinfo {author} {\bibfnamefont {S.-C.}\
  \bibnamefont {Zhang}},\ }\href {\doibase 10.1126/science.1148047} {\bibfield
  {journal} {\bibinfo  {journal} {Science}\ }\textbf {\bibinfo {volume}
  {318}},\ \bibinfo {pages} {766} (\bibinfo {year} {2007})}\BibitemShut
  {NoStop}%
\bibitem [{\citenamefont {Kechedzhi}\ \emph {et~al.}(2008)\citenamefont
  {Kechedzhi}, \citenamefont {Kashuba},\ and\ \citenamefont
  {Fal'ko}}]{Kechedzhi2008}%
  \BibitemOpen
  \bibfield  {author} {\bibinfo {author} {\bibfnamefont {K.}~\bibnamefont
  {Kechedzhi}}, \bibinfo {author} {\bibfnamefont {O.}~\bibnamefont {Kashuba}},
  \ and\ \bibinfo {author} {\bibfnamefont {V.~I.}\ \bibnamefont {Fal'ko}},\
  }\href {\doibase 10.1103/PhysRevB.77.193403} {\bibfield  {journal} {\bibinfo
  {journal} {Phys. Rev. B}\ }\textbf {\bibinfo {volume} {77}},\ \bibinfo
  {pages} {193403} (\bibinfo {year} {2008})}\BibitemShut {NoStop}%
\bibitem [{\citenamefont {Fritz}\ \emph {et~al.}(2008)\citenamefont {Fritz},
  \citenamefont {Schmalian}, \citenamefont {M{\"u}ller},\ and\ \citenamefont
  {Sachdev}}]{Fritz2008}%
  \BibitemOpen
  \bibfield  {author} {\bibinfo {author} {\bibfnamefont {L.}~\bibnamefont
  {Fritz}}, \bibinfo {author} {\bibfnamefont {J.}~\bibnamefont {Schmalian}},
  \bibinfo {author} {\bibfnamefont {M.}~\bibnamefont {M{\"u}ller}}, \ and\
  \bibinfo {author} {\bibfnamefont {S.}~\bibnamefont {Sachdev}},\ }\href
  {\doibase 10.1103/PhysRevB.78.085416} {\bibfield  {journal} {\bibinfo
  {journal} {Phys. Rev. B}\ }\textbf {\bibinfo {volume} {78}},\ \bibinfo
  {pages} {085416} (\bibinfo {year} {2008})}\BibitemShut {NoStop}%
\bibitem [{\citenamefont {Ziegler}(2007)}]{Ziegler2007}%
  \BibitemOpen
  \bibfield  {author} {\bibinfo {author} {\bibfnamefont {K.}~\bibnamefont
  {Ziegler}},\ }\href {\doibase 10.1103/PhysRevB.75.233407} {\bibfield
  {journal} {\bibinfo  {journal} {Phys. Rev. B}\ }\textbf {\bibinfo {volume}
  {75}},\ \bibinfo {pages} {233407} (\bibinfo {year} {2007})}\BibitemShut
  {NoStop}%
\bibitem [{\citenamefont {Kashuba}(2008)}]{Kashuba2008}%
  \BibitemOpen
  \bibfield  {author} {\bibinfo {author} {\bibfnamefont {A.~B.}\ \bibnamefont
  {Kashuba}},\ }\href {\doibase 10.1103/PhysRevB.78.085415} {\bibfield
  {journal} {\bibinfo  {journal} {Phys. Rev. B}\ }\textbf {\bibinfo {volume}
  {78}},\ \bibinfo {pages} {085415} (\bibinfo {year} {2008})}\BibitemShut
  {NoStop}%
\bibitem [{\citenamefont {M\"uller}\ \emph {et~al.}(2009)\citenamefont
  {M\"uller}, \citenamefont {Schmalian},\ and\ \citenamefont
  {Fritz}}]{Muller2009a}%
  \BibitemOpen
  \bibfield  {author} {\bibinfo {author} {\bibfnamefont {M.}~\bibnamefont
  {M\"uller}}, \bibinfo {author} {\bibfnamefont {J.}~\bibnamefont {Schmalian}},
  \ and\ \bibinfo {author} {\bibfnamefont {L.}~\bibnamefont {Fritz}},\ }\href
  {\doibase 10.1103/PhysRevLett.103.025301} {\bibfield  {journal} {\bibinfo
  {journal} {Phys. Rev. Lett.}\ }\textbf {\bibinfo {volume} {103}},\ \bibinfo
  {pages} {025301} (\bibinfo {year} {2009})}\BibitemShut {NoStop}%
\bibitem [{\citenamefont {Briskot}\ \emph {et~al.}(2015)\citenamefont
  {Briskot}, \citenamefont {Sch{\"u}tt}, \citenamefont {Gornyi}, \citenamefont
  {Titov}, \citenamefont {Narozhny},\ and\ \citenamefont
  {Mirlin}}]{Briskot2015}%
  \BibitemOpen
  \bibfield  {author} {\bibinfo {author} {\bibfnamefont {U.}~\bibnamefont
  {Briskot}}, \bibinfo {author} {\bibfnamefont {M.}~\bibnamefont {Sch{\"u}tt}},
  \bibinfo {author} {\bibfnamefont {I.~V.}\ \bibnamefont {Gornyi}}, \bibinfo
  {author} {\bibfnamefont {M.}~\bibnamefont {Titov}}, \bibinfo {author}
  {\bibfnamefont {B.~N.}\ \bibnamefont {Narozhny}}, \ and\ \bibinfo {author}
  {\bibfnamefont {A.~D.}\ \bibnamefont {Mirlin}},\ }\href {\doibase
  10.1103/PhysRevB.92.115426} {\bibfield  {journal} {\bibinfo  {journal} {Phys.
  Rev. B}\ }\textbf {\bibinfo {volume} {92}},\ \bibinfo {pages} {115426}
  (\bibinfo {year} {2015})}\BibitemShut {NoStop}%
\bibitem [{\citenamefont {Narozhny}\ \emph {et~al.}(2015)\citenamefont
  {Narozhny}, \citenamefont {Gornyi}, \citenamefont {Titov}, \citenamefont
  {Sch{\"u}tt},\ and\ \citenamefont {Mirlin}}]{Narozhny2015}%
  \BibitemOpen
  \bibfield  {author} {\bibinfo {author} {\bibfnamefont {B.~N.}\ \bibnamefont
  {Narozhny}}, \bibinfo {author} {\bibfnamefont {I.~V.}\ \bibnamefont
  {Gornyi}}, \bibinfo {author} {\bibfnamefont {M.}~\bibnamefont {Titov}},
  \bibinfo {author} {\bibfnamefont {M.}~\bibnamefont {Sch{\"u}tt}}, \ and\
  \bibinfo {author} {\bibfnamefont {A.~D.}\ \bibnamefont {Mirlin}},\ }\href
  {\doibase 10.1103/PhysRevB.91.035414} {\bibfield  {journal} {\bibinfo
  {journal} {Phys. Rev. B}\ }\textbf {\bibinfo {volume} {91}},\ \bibinfo
  {pages} {035414} (\bibinfo {year} {2015})}\BibitemShut {NoStop}%
\bibitem [{\citenamefont {Lucas}\ \emph {et~al.}(2016)\citenamefont {Lucas},
  \citenamefont {Crossno}, \citenamefont {Fong}, \citenamefont {Kim},\ and\
  \citenamefont {Sachdev}}]{Lucas2016a}%
  \BibitemOpen
  \bibfield  {author} {\bibinfo {author} {\bibfnamefont {A.}~\bibnamefont
  {Lucas}}, \bibinfo {author} {\bibfnamefont {J.}~\bibnamefont {Crossno}},
  \bibinfo {author} {\bibfnamefont {K.~C.}\ \bibnamefont {Fong}}, \bibinfo
  {author} {\bibfnamefont {P.}~\bibnamefont {Kim}}, \ and\ \bibinfo {author}
  {\bibfnamefont {S.}~\bibnamefont {Sachdev}},\ }\href {\doibase
  10.1103/PhysRevB.93.075426} {\bibfield  {journal} {\bibinfo  {journal} {Phys.
  Rev. B}\ }\textbf {\bibinfo {volume} {93}},\ \bibinfo {pages} {075426}
  (\bibinfo {year} {2016})}\BibitemShut {NoStop}%
\bibitem [{\citenamefont {Narozhny}\ \emph {et~al.}(2017)\citenamefont
  {Narozhny}, \citenamefont {Gornyi}, \citenamefont {Mirlin},\ and\
  \citenamefont {Schmalian}}]{Narozhny2017}%
  \BibitemOpen
  \bibfield  {author} {\bibinfo {author} {\bibfnamefont {B.~N.}\ \bibnamefont
  {Narozhny}}, \bibinfo {author} {\bibfnamefont {I.~V.}\ \bibnamefont
  {Gornyi}}, \bibinfo {author} {\bibfnamefont {A.~D.}\ \bibnamefont {Mirlin}},
  \ and\ \bibinfo {author} {\bibfnamefont {J.}~\bibnamefont {Schmalian}},\
  }\href {\doibase 10.1002/andp.201700043} {\bibfield  {journal} {\bibinfo
  {journal} {Annalen der Physik}\ }\textbf {\bibinfo {volume} {529}},\ \bibinfo
  {pages} {1700043} (\bibinfo {year} {2017})}\BibitemShut {NoStop}%
\bibitem [{\citenamefont {Lucas}\ and\ \citenamefont {Fong}(2018)}]{Lucas2018}%
  \BibitemOpen
  \bibfield  {author} {\bibinfo {author} {\bibfnamefont {A.}~\bibnamefont
  {Lucas}}\ and\ \bibinfo {author} {\bibfnamefont {K.~C.}\ \bibnamefont
  {Fong}},\ }\href {\doibase 10.1088/1361-648X/aaa274} {\bibfield  {journal}
  {\bibinfo  {journal} {Journal of Physics: Condensed Matter}\ }\textbf
  {\bibinfo {volume} {30}},\ \bibinfo {pages} {053001} (\bibinfo {year}
  {2018})}\BibitemShut {NoStop}%
\bibitem [{Note1()}]{Note1}%
  \BibitemOpen
  \bibinfo {note} {The full Hamiltonian including Coulomb interaction is not
  relativistic i.e.\ not invariant with respect to Lorentz transformations.
  However, the word ``relativistic'' is widely used in the modern literature of
  condensed matter physics (e.g. on graphene or surface states of 3D
  topological insulators) referring to the Dirac kinetic part of the
  Hamiltonian.}\BibitemShut {Stop}%
\bibitem [{\citenamefont {Seo}\ \emph {et~al.}(2017)\citenamefont {Seo},
  \citenamefont {Song}, \citenamefont {Kim}, \citenamefont {Sachdev},\ and\
  \citenamefont {Sin}}]{Sachdev2017}%
  \BibitemOpen
  \bibfield  {author} {\bibinfo {author} {\bibfnamefont {Y.}~\bibnamefont
  {Seo}}, \bibinfo {author} {\bibfnamefont {G.}~\bibnamefont {Song}}, \bibinfo
  {author} {\bibfnamefont {P.}~\bibnamefont {Kim}}, \bibinfo {author}
  {\bibfnamefont {S.}~\bibnamefont {Sachdev}}, \ and\ \bibinfo {author}
  {\bibfnamefont {S.-J.}\ \bibnamefont {Sin}},\ }\href {\doibase
  10.1103/PhysRevLett.118.036601} {\bibfield  {journal} {\bibinfo  {journal}
  {Phys. Rev. Lett.}\ }\textbf {\bibinfo {volume} {118}},\ \bibinfo {pages}
  {036601} (\bibinfo {year} {2017})}\BibitemShut {NoStop}%
\bibitem [{\citenamefont {Lifshitz}\ and\ \citenamefont
  {Pitaevskii}(1981)}]{ll10}%
  \BibitemOpen
  \bibfield  {author} {\bibinfo {author} {\bibfnamefont {E.~M.}\ \bibnamefont
  {Lifshitz}}\ and\ \bibinfo {author} {\bibfnamefont {L.~P.}\ \bibnamefont
  {Pitaevskii}},\ }\href@noop {} {\emph {\bibinfo {title} {Physical
  Kinetics}}}\ (\bibinfo  {publisher} {Pergamon Press, Oxford},\ \bibinfo
  {year} {1981})\BibitemShut {NoStop}%
\bibitem [{\citenamefont {van Ostaay}\ \emph {et~al.}(2011)\citenamefont {van
  Ostaay}, \citenamefont {Akhmerov}, \citenamefont {Beenakker},\ and\
  \citenamefont {Wimmer}}]{Ostaay2011}%
  \BibitemOpen
  \bibfield  {author} {\bibinfo {author} {\bibfnamefont {J.~A.~M.}\
  \bibnamefont {van Ostaay}}, \bibinfo {author} {\bibfnamefont {A.~R.}\
  \bibnamefont {Akhmerov}}, \bibinfo {author} {\bibfnamefont {C.~W.~J.}\
  \bibnamefont {Beenakker}}, \ and\ \bibinfo {author} {\bibfnamefont
  {M.}~\bibnamefont {Wimmer}},\ }\href {\doibase 10.1103/PhysRevB.84.195434}
  {\bibfield  {journal} {\bibinfo  {journal} {Phys. Rev. B}\ }\textbf {\bibinfo
  {volume} {84}},\ \bibinfo {pages} {195434} (\bibinfo {year}
  {2011})}\BibitemShut {NoStop}%
\bibitem [{\citenamefont {Kharitonov}\ \emph {et~al.}(2017)\citenamefont
  {Kharitonov}, \citenamefont {Mayer},\ and\ \citenamefont
  {Hankiewicz}}]{Kharitonov2017}%
  \BibitemOpen
  \bibfield  {author} {\bibinfo {author} {\bibfnamefont {M.}~\bibnamefont
  {Kharitonov}}, \bibinfo {author} {\bibfnamefont {J.-B.}\ \bibnamefont
  {Mayer}}, \ and\ \bibinfo {author} {\bibfnamefont {E.~M.}\ \bibnamefont
  {Hankiewicz}},\ }\href {\doibase 10.1103/PhysRevLett.119.266402} {\bibfield
  {journal} {\bibinfo  {journal} {Phys. Rev. Lett.}\ }\textbf {\bibinfo
  {volume} {119}},\ \bibinfo {pages} {266402} (\bibinfo {year}
  {2017})}\BibitemShut {NoStop}%
\bibitem [{\citenamefont {Fuchs}(1938)}]{Fuchs1938}%
  \BibitemOpen
  \bibfield  {author} {\bibinfo {author} {\bibfnamefont {K.}~\bibnamefont
  {Fuchs}},\ }\href@noop {} {\bibfield  {journal} {\bibinfo  {journal}
  {Proceedings of the Cambridge Philosophical Society. Mathematical and
  Physical Sciences}\ }\textbf {\bibinfo {volume} {34}},\ \bibinfo {pages}
  {100} (\bibinfo {year} {1938})}\BibitemShut {NoStop}%
\bibitem [{\citenamefont {Sondheimer}(1952)}]{Sondheimer1952}%
  \BibitemOpen
  \bibfield  {author} {\bibinfo {author} {\bibfnamefont {E.~H.}\ \bibnamefont
  {Sondheimer}},\ }\bibfield  {booktitle} {\emph {\bibinfo {booktitle}
  {Advances in Physics}},\ }\href {\doibase 10.1080/00018735200101151}
  {\bibfield  {journal} {\bibinfo  {journal} {Advances in Physics}\ }\textbf
  {\bibinfo {volume} {1}},\ \bibinfo {pages} {1} (\bibinfo {year}
  {1952})}\BibitemShut {NoStop}%
\bibitem [{\citenamefont {Knudsen}(1910)}]{Knudsen1910}%
  \BibitemOpen
  \bibfield  {author} {\bibinfo {author} {\bibfnamefont {M.}~\bibnamefont
  {Knudsen}},\ }\href {\doibase 10.1002/andp.19103360310} {\bibfield  {journal}
  {\bibinfo  {journal} {Ann. Phys.}\ }\textbf {\bibinfo {volume} {336}},\
  \bibinfo {pages} {633} (\bibinfo {year} {1910})}\BibitemShut {NoStop}%
\bibitem [{\citenamefont {Ostrovsky}\ \emph {et~al.}(2006)\citenamefont
  {Ostrovsky}, \citenamefont {Gornyi},\ and\ \citenamefont
  {Mirlin}}]{Mirlin2006}%
  \BibitemOpen
  \bibfield  {author} {\bibinfo {author} {\bibfnamefont {P.~M.}\ \bibnamefont
  {Ostrovsky}}, \bibinfo {author} {\bibfnamefont {I.~V.}\ \bibnamefont
  {Gornyi}}, \ and\ \bibinfo {author} {\bibfnamefont {A.~D.}\ \bibnamefont
  {Mirlin}},\ }\href {\doibase 10.1103/PhysRevB.74.235443} {\bibfield
  {journal} {\bibinfo  {journal} {Phys. Rev. B}\ }\textbf {\bibinfo {volume}
  {74}},\ \bibinfo {pages} {235443} (\bibinfo {year} {2006})}\BibitemShut
  {NoStop}%
\bibitem [{\citenamefont {Ostrovsky}\ \emph {et~al.}(2007)\citenamefont
  {Ostrovsky}, \citenamefont {Gornyi},\ and\ \citenamefont
  {Mirlin}}]{Mirlin2007}%
  \BibitemOpen
  \bibfield  {author} {\bibinfo {author} {\bibfnamefont {P.~M.}\ \bibnamefont
  {Ostrovsky}}, \bibinfo {author} {\bibfnamefont {I.~V.}\ \bibnamefont
  {Gornyi}}, \ and\ \bibinfo {author} {\bibfnamefont {A.~D.}\ \bibnamefont
  {Mirlin}},\ }\href {\doibase 10.1103/PhysRevLett.98.256801} {\bibfield
  {journal} {\bibinfo  {journal} {Phys. Rev. Lett.}\ }\textbf {\bibinfo
  {volume} {98}},\ \bibinfo {pages} {256801} (\bibinfo {year}
  {2007})}\BibitemShut {NoStop}%
\bibitem [{\citenamefont {Callaway}(1959)}]{Callaway1959}%
  \BibitemOpen
  \bibfield  {author} {\bibinfo {author} {\bibfnamefont {J.}~\bibnamefont
  {Callaway}},\ }\href {\doibase 10.1103/PhysRev.113.1046} {\bibfield
  {journal} {\bibinfo  {journal} {Phys. Rev.}\ }\textbf {\bibinfo {volume}
  {113}},\ \bibinfo {pages} {1046} (\bibinfo {year} {1959})}\BibitemShut
  {NoStop}%
\bibitem [{\citenamefont {Gennaro}\ and\ \citenamefont
  {Rettori}(1984)}]{Gennaro1984a}%
  \BibitemOpen
  \bibfield  {author} {\bibinfo {author} {\bibfnamefont {S.~D.}\ \bibnamefont
  {Gennaro}}\ and\ \bibinfo {author} {\bibfnamefont {A.}~\bibnamefont
  {Rettori}},\ }\href {http://stacks.iop.org/0305-4608/14/i=12/a=001}
  {\bibfield  {journal} {\bibinfo  {journal} {Journal of Physics F: Metal
  Physics}\ }\textbf {\bibinfo {volume} {14}},\ \bibinfo {pages} {L237}
  (\bibinfo {year} {1984})}\BibitemShut {NoStop}%
\bibitem [{\citenamefont {Gennaro}\ and\ \citenamefont
  {Rettori}(1985)}]{Gennaro1985a}%
  \BibitemOpen
  \bibfield  {author} {\bibinfo {author} {\bibfnamefont {S.~D.}\ \bibnamefont
  {Gennaro}}\ and\ \bibinfo {author} {\bibfnamefont {A.}~\bibnamefont
  {Rettori}},\ }\href {http://stacks.iop.org/0305-4608/15/i=10/a=013}
  {\bibfield  {journal} {\bibinfo  {journal} {Journal of Physics F: Metal
  Physics}\ }\textbf {\bibinfo {volume} {15}},\ \bibinfo {pages} {2177}
  (\bibinfo {year} {1985})}\BibitemShut {NoStop}%
\bibitem [{\citenamefont {Poiseuille}(1838)}]{Poiseuille1838}%
  \BibitemOpen
  \bibfield  {author} {\bibinfo {author} {\bibfnamefont {J.~L.~M.}\
  \bibnamefont {Poiseuille}},\ }\href@noop {} {\bibfield  {journal} {\bibinfo
  {journal} {Extraits des Proces-Verbaux des Seances Pendant}\ }\textbf
  {\bibinfo {volume} {3}},\ \bibinfo {pages} {77} (\bibinfo {year}
  {1838})}\BibitemShut {NoStop}%
\bibitem [{\citenamefont {M\"uller}\ \emph {et~al.}(2008)\citenamefont
  {M\"uller}, \citenamefont {Fritz},\ and\ \citenamefont
  {Sachdev}}]{Muller2008b}%
  \BibitemOpen
  \bibfield  {author} {\bibinfo {author} {\bibfnamefont {M.}~\bibnamefont
  {M\"uller}}, \bibinfo {author} {\bibfnamefont {L.}~\bibnamefont {Fritz}}, \
  and\ \bibinfo {author} {\bibfnamefont {S.}~\bibnamefont {Sachdev}},\ }\href
  {\doibase 10.1103/PhysRevB.78.115406} {\bibfield  {journal} {\bibinfo
  {journal} {Phys. Rev. B}\ }\textbf {\bibinfo {volume} {78}},\ \bibinfo
  {pages} {115406} (\bibinfo {year} {2008})}\BibitemShut {NoStop}%
\bibitem [{\citenamefont {Siegel}\ \emph {et~al.}(2013)\citenamefont {Siegel},
  \citenamefont {Regan}, \citenamefont {Fedorov}, \citenamefont {Zettl},\ and\
  \citenamefont {Lanzara}}]{Siegel2013}%
  \BibitemOpen
  \bibfield  {author} {\bibinfo {author} {\bibfnamefont {D.~A.}\ \bibnamefont
  {Siegel}}, \bibinfo {author} {\bibfnamefont {W.}~\bibnamefont {Regan}},
  \bibinfo {author} {\bibfnamefont {A.~V.}\ \bibnamefont {Fedorov}}, \bibinfo
  {author} {\bibfnamefont {A.}~\bibnamefont {Zettl}}, \ and\ \bibinfo {author}
  {\bibfnamefont {A.}~\bibnamefont {Lanzara}},\ }\href {\doibase
  10.1103/PhysRevLett.110.146802} {\bibfield  {journal} {\bibinfo  {journal}
  {Phys. Rev. Lett.}\ }\textbf {\bibinfo {volume} {110}},\ \bibinfo {pages}
  {146802} (\bibinfo {year} {2013})}\BibitemShut {NoStop}%
\bibitem [{\citenamefont {Foster}\ and\ \citenamefont
  {Aleiner}(2008)}]{Foster2008}%
  \BibitemOpen
  \bibfield  {author} {\bibinfo {author} {\bibfnamefont {M.~S.}\ \bibnamefont
  {Foster}}\ and\ \bibinfo {author} {\bibfnamefont {I.~L.}\ \bibnamefont
  {Aleiner}},\ }\href {\doibase 10.1103/PhysRevB.77.195413} {\bibfield
  {journal} {\bibinfo  {journal} {Phys. Rev. B}\ }\textbf {\bibinfo {volume}
  {77}},\ \bibinfo {pages} {195413} (\bibinfo {year} {2008})}\BibitemShut
  {NoStop}%
\bibitem [{\citenamefont {Sheehy}\ and\ \citenamefont
  {Schmalian}(2007)}]{Sheehy2007}%
  \BibitemOpen
  \bibfield  {author} {\bibinfo {author} {\bibfnamefont {D.~E.}\ \bibnamefont
  {Sheehy}}\ and\ \bibinfo {author} {\bibfnamefont {J.}~\bibnamefont
  {Schmalian}},\ }\href {\doibase 10.1103/PhysRevLett.99.226803} {\bibfield
  {journal} {\bibinfo  {journal} {Phys. Rev. Lett.}\ }\textbf {\bibinfo
  {volume} {99}},\ \bibinfo {pages} {226803} (\bibinfo {year}
  {2007})}\BibitemShut {NoStop}%
\bibitem [{\citenamefont {Kotov}\ \emph {et~al.}(2012)\citenamefont {Kotov},
  \citenamefont {Uchoa}, \citenamefont {Pereira}, \citenamefont {Guinea},\ and\
  \citenamefont {Castro~Neto}}]{Kotov2012}%
  \BibitemOpen
  \bibfield  {author} {\bibinfo {author} {\bibfnamefont {V.~N.}\ \bibnamefont
  {Kotov}}, \bibinfo {author} {\bibfnamefont {B.}~\bibnamefont {Uchoa}},
  \bibinfo {author} {\bibfnamefont {V.~M.}\ \bibnamefont {Pereira}}, \bibinfo
  {author} {\bibfnamefont {F.}~\bibnamefont {Guinea}}, \ and\ \bibinfo {author}
  {\bibfnamefont {A.~H.}\ \bibnamefont {Castro~Neto}},\ }\href {\doibase
  10.1103/RevModPhys.84.1067} {\bibfield  {journal} {\bibinfo  {journal} {Rev.
  Mod. Phys.}\ }\textbf {\bibinfo {volume} {84}},\ \bibinfo {pages} {1067}
  (\bibinfo {year} {2012})}\BibitemShut {NoStop}%
\bibitem [{\citenamefont {Berry}\ and\ \citenamefont
  {Mondragon}(1987)}]{Berry1987}%
  \BibitemOpen
  \bibfield  {author} {\bibinfo {author} {\bibfnamefont {M.~V.}\ \bibnamefont
  {Berry}}\ and\ \bibinfo {author} {\bibfnamefont {R.~J.}\ \bibnamefont
  {Mondragon}},\ }\href {\doibase 10.2307/2398295} {\bibfield  {journal}
  {\bibinfo  {journal} {Proceedings of the Royal Society of London. Series A,
  Mathematical and Physical Sciences}\ }\textbf {\bibinfo {volume} {412}},\
  \bibinfo {pages} {53} (\bibinfo {year} {1987})}\BibitemShut {NoStop}%
\bibitem [{\citenamefont {Torre}\ \emph {et~al.}(2015)\citenamefont {Torre},
  \citenamefont {Tomadin}, \citenamefont {Geim},\ and\ \citenamefont
  {Polini}}]{Torre2015}%
  \BibitemOpen
  \bibfield  {author} {\bibinfo {author} {\bibfnamefont {I.}~\bibnamefont
  {Torre}}, \bibinfo {author} {\bibfnamefont {A.}~\bibnamefont {Tomadin}},
  \bibinfo {author} {\bibfnamefont {A.~K.}\ \bibnamefont {Geim}}, \ and\
  \bibinfo {author} {\bibfnamefont {M.}~\bibnamefont {Polini}},\ }\href
  {\doibase 10.1103/PhysRevB.92.165433} {\bibfield  {journal} {\bibinfo
  {journal} {Phys. Rev. B}\ }\textbf {\bibinfo {volume} {92}},\ \bibinfo
  {pages} {165433} (\bibinfo {year} {2015})}\BibitemShut {NoStop}%
\end{thebibliography}%

\end{document}